\documentclass{aa} % for a referee version
\usepackage{graphicx}
\usepackage{natbib}
\bibpunct{(}{)}{;}{a}{}{,}
\usepackage{longtable,lscape}
\usepackage{txfonts}
\newcommand{\kms}{~km~s$^{-1}$}
\newcommand{\teff}{$T_{\rm eff}$}
\newcommand{\logg}{$\log g$}

\newcommand{\vt}{$v_{\rm micro}$}

\newcommand{\ebv}{$E(B-V)$}
\begin{document}
   \title{Chemical Composition of Extremely Metal-Poor Stars in the
Sextans Dwarf Spheroidal Galaxy.\thanks{Based on data collected at
Subaru Telescope, which is operated by the National Astronomical
Observatory of Japan. }}

%   \subtitle{I. Overviewing the $\kappa$-mechanism}

   \author{W. Aoki
          \inst{1,2}
          \and
          N. Arimoto\inst{1,2}
          \and
          K. Sadakane\inst{3}
          \and
          E. Tolstoy\inst{4}
          \and
          G. Battaglia\inst{5}
          \and
          P. Jablonka\inst{6}
          \and
          M. Shetrone\inst{7}
          \and
          B. Letarte\inst{8}
          \and
          M. Irwin\inst{9}
          \and
          V. Hill\inst{10}
          \and
          P. Francois\inst{11}
          \and
          K. Venn\inst{12}
          \and
          F. Primas\inst{5}
          \and
          A. Helmi\inst{4}
          \and
          A. Kaufer\inst{13}
          \and
          M. Tafelmeyer\inst{6}
          \and
          T. Szeifert\inst{13}
          \and
          C. Babusiaux\inst{10}
          }
%          N. Arimoto\inst{1,2}\fnmsep\thanks{Just to show the usage
%          of the elements in the author field}

   \institute{National Astronomical Observatory of Japan, Mitaka, Tokyo 181-8588, Japan 
              \email{aoki.wako@nao.ac.jp, arimoto.n@nao.ac.jp}
         \and
             Department of Astronomical Science, Graduate University of Advanced
  Studies, Mitaka, Tokyo 181-8588, Japan 
%             \email{}
         \and
         Astronomical Institute, Osaka Kyoiku University, Asahigaoka,
         Kashiwara, Osaka 582-8582, Japan 
             \email{sadakane@cc.osaka-kyoiku.ac.jp}
         \and
         Kapteyn Astronomical Institute, University of Groningen,
         P.O. Box 800, 9700 AV Groningen, Netherlands
             \email{etolstoy@astro.rug.nl, ahelmi@astro.rug.nl}
         \and
         European Southern Observatory, Karl-Schwarzschild-Strasse 2,
         85748 Garching bei M\"{u}nchen, Germany 
             \email{gbattagl@eso.org, fprimas@eso.org}
         \and
         Observatoire de Gen\'{e}ve, Laboratoire d'Astrophysique de
         l'Ecole Polytechnique F\'{e}d\'{e}rale de Lausanne (EPFL), CH-1290
         Sauverny, Switzerland 
             \email{Pascale.Jablonka@obs.unige.ch, martin.tafelmeyer@epfl.ch}
         \and
         University of Texas, McDonald Observatory, HC75 Box 1337-McD, Fort Davis, TX 79734, USA 
             \email{shetrone@astro.as.utexas.edu}
         \and
         California Institute of Technology, Pasadena, CA 91125, USA 
             \email{bruno@astro.caltech.edu}
         \and 
         Institute of Astronomy, Madingley Road, Cambridge CB03 0HA, UK
             \email{mike@ast.cam.ac.uk}
         \and
         GEPI, Observatoire de Paris, CNRS, Universit\'{e} Paris Diderot,
         Place Jules Janssen 92190 Meudon, France
             \email{Vanessa.Hill@obspm.fr, carine.babusiaux@obspm.fr}
         \and 
         Observatoire de Paris-Meudon, GEPI, 61 avenue de l'Observatoire, 75014 Paris, France 
             \email{patrick.Francois@obspm.fr}
         \and
         Department of Physics and Astronomy, University of Victoria,
         Elliott Building, 3800 Finnerty Road, Victoria, BC V8P 5C2,
         Canada 
             \email{kvenn@uvic.ca}
         \and
         European Southern Observatory, Alonso de C\'{o}rdova 3107, Santiago, Chile 
             \email{akaufer@eso.org, tszeifer@eso.org}
                     }
   \date{Received; accepted}

% \abstract{}{}{}{}{} 
% 5 {} token are mandatory
 
  \abstract
  % context heading (optional)
  % {} leave it empty if necessary  
   {Individual stars in dwarf spheroidal galaxies around the Milky 
Way Galaxy have been studied both photometrically and spectroscopically.
Extremely metal-poor stars among them are very valuable
because they should record the early enrichment in the Local Group. However,
our understanding of these stars is very limited because detailed
chemical abundance measurements are needed from high resolution spectroscopy.}
  % aims heading (mandatory)
   {To constrain the formation and chemical evolution of dwarf galaxies, 
metallicity and chemical composition of extremely metal-poor stars are investigated.}
  % methods heading (mandatory)
   {Chemical abundances of six extremely metal-poor ([Fe/H]$<-2.5$)
     stars in the Sextans dwarf spheroidal galaxy are determined based
     on high resolution spectroscopy ($R=40,000$) with the Subaru
     Telescope High Dispersion Spectrograph.}
  % results heading (mandatory)
   {(1) The Fe abundances derived from the high resolution spectra
     are in good agreement with the metallicity estimated from the Ca
     triplet lines in low resolution spectra. The lack of stars
     with [Fe/H]$\lesssim-3$ in Sextans, found by previous estimates from 
     the Ca triplet, is confirmed by our measurements,  although we
     note that high resolution spectroscopy for a larger sample of stars 
     will be necessary to estimate the true fraction of stars with
     such low metallicity.
    (2) While one object shows an overabundance of Mg (similar to Galactic 
     halo stars), the Mg/Fe ratios of the remaining five stars are similar 
     to the solar value.  This is the first time that low Mg/Fe ratios at 
     such low metallicities have been found in a dwarf spheroidal galaxy.  
     No evidence for over-abundances of Ca and Ti are found in these five 
     stars, though the measurements for these elements are less certain. 
     Possible mechanisms to produce low Mg/Fe ratios, with respect to that 
     of Galactic halo stars, are discussed. 
    (3) Ba is under-abundant in four objects, while the remaining two stars 
     exhibit large and moderate
     excesses of this element. The abundance distribution of Ba in
     this galaxy is similar to that in the Galactic halo, indicating
     that the enrichment of heavy elements, probably by the r-process,
     started at metallicities [Fe/H] $\le -2.5$, as found in the Galactic halo.}
% conclusions heading (optional), leave it empty if necessary
   {}


   \maketitle
%
%________________________________________________________________

\section{Introduction}

The Local Group includes a number of dwarf galaxies that have been
evolving while interacting with the giant spiral galaxies, the Milky Way and
M31. Among these dwarf galaxies, the dwarf spheroidals contain little gas and
dust, and are quiescent at present. They are located close to the giant spirals, 
compared to gas-rich dwarf irregulars, providing a unique opportunity to study 
these galaxies in detail, based on photometry and spectroscopy of individual 
stars.

Previous high resolution spectroscopic studies of bright red giants in
several dwarf spheroidals \citep{shetrone01, shetrone03} revealed that
the abundance ratios between $\alpha$-elements and Fe-peak elements
(e.g., Mg/Fe) decreases with increasing metallicity even in metal-poor
stars ($-2<$[Fe/H]$<-1$)\footnote{[A/B] = $\log(N_{\rm A}/N_{\rm B})-
  \log(N_{\rm A}/N_{\rm B})_{\odot}$, and $\log\epsilon_{\rm A} =
  \log(N_{\rm A}/N_{\rm H})+12$ for elements A and B.}. This is quite
different from the trend found in the Galactic halo, where metal poor
stars show an almost constant enhancement in the $\alpha$/Fe ratio,
with only a few exceptions having significantly higher or lower
values.  Moreover, the trend in the $\alpha$/Fe ratios, as a function
of metallicity, varies between the individual dwarf galaxies as
well. Differences of abundance ratios are also seen in neutron-capture
elements (e.g. Ba/Fe, Ba/Y) between stars in dwarf spheroidal galaxies
and those in field stars \citet[e.g., ][]{venn04}.

Another difference found between dwarf spheroidals and the Galactic
halo is the metallicity distribution, in particular that of the low
metallicity end. \citet{helmi06} showed the deficiency of stars with 
[Fe/H]$<-3$ in four dwarf spheroidals based on the measurements of the Ca
{\small II} triplet in low resolution spectra obtained
with the VLT/FLAMES. This difference, as well as the difference in the
$\alpha$/Fe ratios, suggests that these dwarf spheroidals are not the
direct relics of the building blocks of the Galactic halo. However,
the metallicity distribution in the extremely low metallicity stars 
from \citet{helmi06} require confirmation from high resolution spectroscopy. 
This is because the metallicities determined from the Ca  
triplet method have only been calibrated for stars with [Fe/H]$>-2.5$, 
i.e., the metallicities of globular cluster stars used in the calibration. 
Recently, \citet{battaglia08} confirmed that the metallicity from the 
Ca triplet is in good agreement with the Fe abundances based 
on high resolution spectra for the 129 stars in the Sculptor and Fornax 
dwarf spheroidal galaxies, however those metallicities were still 
[Fe/H]$>-2.5$, thus further studies of lower metallicity stars are required. 

Observational studies of extremely metal-poor ([Fe/H]$<-2.5$) stars in
dwarf spheroidal galaxies based on high resolution spectroscopy are still 
very limited. Detailed chemical abundance analyses have been published 
for only three red giants in three galaxies to date (Sextans, Draco and
Ursa Minor) by \citet{shetrone01}, \citet{fulbright04} and
\citet{sadakane04}. These three stars show quite interesting abundance
patterns, e.g.,  a deficiency of Na and neutron-capture elements. 
However, since only one object has been studied in each galaxy, then
the early chemical enrichment of dwarf spheroidals is still far from
conclusive.

 Quite recently, extremely metal-poor stars have been found in
  ultra-faint dwarf galaxies, which have been discovered with SDSS
  \citep{kirby08, norris08}. High resolution spectroscopy has been
  also made for several stars in these galaxies
  \citep{koch08,frebel09}. Comparisons of extremely metal-poor stars
  in classical dwarf spheroidals and those in ultra-faint dwarf
  galaxies are useful for understanding of their formation and
  evolution.

To investigate the chemical nature of the lowest metallicity stars in 
dwarf galaxies, we have gathered high resolution spectra of individual 
stars in dwarf galaxies using the Subaru Telescope and VLT. 
In this paper, we report on the chemical composition of six extremely 
metal-poor stars in the Sextans dwarf spheroidal galaxy from observations
taken with the Subaru High Dispersion Spectrograph.  \citet{helmi06} 
suggest that this galaxy includes a relatively large number of stars with 
[Fe/H] $<-2.5$.  The analysis presented here provides Fe abundances to 
examine the calibration of the metallicity from the Ca triplet
in more metal-poor stars, as well as new abundance ratios of 
$\alpha$-elements, other Fe-peak elements, and neutron-capture elements.

%__________________________________________________________________

\section{Observation and Measurements}

\subsection{Sample Selection and Photometric Data}

We selected stars whose metallicity ([Fe/H]) was estimated to be lower
than $-2.5$ by \citet{helmi06} based on the VLT/FLAMES low
resolution spectroscopy for the Ca triplet region.  The list of
objects with their coordinates is given in Table~\ref{tab:obj}. The
table includes the two stars S~10--12 and S~11--36, which were not
analyzed in the present work, because the former has only very broad 
absorption features with a systemic velocity close to 0~\kms, and the 
latter is a carbon-rich star.\footnote{Although the data quality is 
  insufficient for the present purpose, our analysis of the spectrum 
  of S~11--36 indicates that this star is extremely metal-poor
  ([Fe/H]=$-2.9\pm0.3$) and carbon-rich ([C/Fe]=$+1.9\pm 0.3$). The
  effective temperature of 4500~K is estimated from the $V-K$ color,
  while $\log g =1.0$ and {\vt}=$2.3$~{\kms} have been assumed in the
  analysis based on the typical values we found for the Sextans giants
  (see \S~3 for the details of our analysis technique). 
  The carbon abundance is estimated from the C$_{2}$ Swan band at 
  5165~{\AA}. Ba shows some excess ([Ba/Fe]$=+0.8$). Such a moderate 
  excess of Ba is found in carbon-enhanced extremely metal-poor stars 
  in the Galactic halo (e.g. Aoki et al. 2007a), and suggests that 
  the origin of the carbon-excess is in the nucleosynthesis of an AGB 
  star. However,
  given the fact that the non-carbon-rich star S~15--19 in the Sextans
  dwarf galaxy also shows a moderate excess of Ba (\S~3), the origins
  of the carbon and Ba in S~11-36 are not definitively determined.}
 
%The optical photometry ($V$ and $I$ of the standard Johnson-Cousins
%system) was taken with WFC on the INT on La Palma and WFI on the
%ESO/2.2m telescope, while the near-infrared $JHK$ data are from WFCAM
%observations on UKIRT on Mauna Kea.  The photometric data for $V, I$
%and $K$ are listed in Table~\ref{tab:param}.  Typical photometric
%errors are 0.01-0.02 magnitudes for $V$ and $I$, and 0.02-0.03
%magnitudes for $K$. The foreground reddening of $E(B-V)=0.038$ is
%estimated from the dust maps of \citet{schlegel98}. The value agrees
%with the estimate of reddening listed by \citet{mateo98}:
%$E(B-V)=0.03\pm 0.01$.  The extinction in each photometric band is
%derived using the reddening relations in Table 6 of
%\citet{schlegel98}.

The optical photometry for the central regions was taken with the WFC on
the Isaac Newton Telescope on La Palma and for the outer regions with WFI
on the ESO/2.2m telescope on La Silla, while the near-infrared $JHK$
observations were all taken with WFCAM on UKIRT on Mauna Kea as part of the
UKIDSS surveys \citep{lawrence07}.   Optical data were processed and
calibrated using variants of the data reduction pipeline developed for the
INT Wide Field Survey \footnote{http://www.ast.cam.ac.uk/~wfcsur/} and the
near-infrared $JHK$ data was processed using the VDFS pipeline developed
for WFCAM and VISTA surveys (see Irwin \& Lewis 2001; and Irwin et al. 2004
for general details of the pipeline products and procedures).  The optical
data in the instrumental $V$ and $i$ passbands was converted to $V$ and $I$
on the standard Johnson-Cousins system using previously derived color
equations for these camera systems.  Near-infrared calibration of WFCAM
data is based on 2MASS and is on the MKO system \citep{hodgkin09}.
Photometric data for $V, I$ and $K$ are listed in Table~\ref{tab:param} and
have typical errors of 0.01-0.02 magnitudes for $V$ and $I$, and 0.02-0.03
magnitudes for $K$. The foreground reddening of $E(B-V)=0.038$ is
estimated from the dust maps of \citet{schlegel98}. The value agrees
with the estimate of reddening listed by \citet{mateo98}:
$E(B-V)=0.03\pm 0.01$.  The extinction in each photometric band is
derived using the reddening relations in Table 6 of
\citet{schlegel98}.

The extremely metal-poor, halo red giant, HD~88609 was observed as a
comparison star.  Its optical photometric data were taken from
\citet{carney83}, and the $K$ magnitude from the 2MASS catalogue
\citep{skrutskie06}.  The $V-I$ color in Table~\ref{tab:param} for
HD~88609 is transformed from the Johnson $I$ magnitude from
\citet{carney83} to the Kron-Cousins system.

\subsection{High Resolution Spectroscopy}

High resolution spectra of extremely metal-poor star candidates in
Sextans were obtained with the Subaru Telescope High Dispersion
Spectrograph (HDS, Noguchi et al. 2002) in May 2005 and January 2007.
The spectra cover 4400--7100~{\AA} with a resolving power of 40,000.
The exposure time and the signal-to-noise (S/N) ratios per
1.8~km$^{-1}$ pixel at 5180~{\AA} are given in
Table~\ref{tab:obj}. The seeing conditions varied from 0.6 to 1.4 arcsec
during the January 2007 run, and 0.6 to 1.1 arcsec during the May 2005 
observations.

Standard data reduction procedures were carried out with the IRAF echelle
package\footnote{IRAF is distributed by the National Optical Astronomy
Observatories, which is operated by the Association of Universities
for Research in Astronomy, Inc. under cooperative agreement with the
National Science Foundation.}.   Cosmic ray strikes were 
removed using the procedures described by \citet{aoki05}. The sky
background is significant in some cases because those observations 
were made close to the full moon.  We attempted to estimate the 
sky background for each exposure, and excluded those exposures where
the background dominates the object's light, or when 
the estimate is uncertain due to the poor seeing conditions.
The sky background was removed in the remaining frames. Individual
spectra were combined after the wavelength calibration. 
The quality of the sky subtraction was examined around the 
Mg {\small I} b lines (5160--5190~{\AA}), where the broad triplet 
features of the solar spectrum can be seen if the sky subtraction is 
not sufficient.  The exposure times and S/N ratios in Table~\ref{tab:obj} 
represent the final spectra, after the selection of exposures 
and the sky subtraction.

\subsection{Measurements of Equivalent Widths and Radial Velocities}

We adopted the line list compiled by Aoki et al. (2009, in
preparation) for studies of extremely metal-poor stars in the Galactic
halo. Equivalent widths for isolated absorption lines are measured by
fitting Gaussian profiles.  This technique could underestimate the
  equivalent widths for Na I D lines and Mg I b lines, which are the
  strongest among the lines studied here. However, we applied the
  spectrum sysnthesis for these lines and found that the wing
  component of these lines is not evident in our stars at the
  resolution and S/N of our spectra.

The radial velocity of each object is measured from clean Fe lines
through the above Gaussian fitting procedure. The heliocentric radial
velocity is given in Table~\ref{tab:obj}. The random error of the
measurement given in the table is estimated as $\sigma_{v} N^{-1/2}$,
where $\sigma_{v}$ is the standard deviation of the values derived
from individual spectral lines and $N$ is the number of lines
used. The results agree well with the Ca triplet measurements from the
VLT/FLAMES data.  The only exception is S~12--28, where our result is
14~{\kms} smaller than that from the Ca triplet lines.  This
disagreement is within the 3$\sigma$ measurement error on the Ca triplet
lines, therefore further measurements are required before concluding
that this star belongs to a binary system.

\section{Chemical Abundance Analyses}

\subsection{Stellar parameters}

The effective temperature ({\teff}) is estimated from the $(V-K)_{0}$ and
$(V-I)_{0}$ color indicies using the temperature scale of
\citet{alonso99}. We assumed [Fe/H]$=-3.0$ for the calculation. 
The {\teff} results for HD~88609 from the two colors agree very
well, however {\teff} from $V-I$ for the Sextans objects is systematically
higher by 150~K, on average, than that from $V-K$. This discrepancy is
not explained by a small increase in the reddening correction. 
Although the reason for this discrepancy is unclear, the
values from the $V-K$ are adopted in this analysis.  Our 
estimated uncertainty in the effective temperatures is $\pm$150~K.

Other stellar parameters are determined through the standard LTE
analysis of the Fe {\small I} and Fe {\small II} lines, using the
model atmospheres from \citet{kurucz93}. We adopted the grid of models
calculated with the new opacity distribution functions and
assuming no convective overshoot \citep{castelli03}. The
microturbulence ({\vt}) and gravity ($g$ in cgs unit) are determined
such that the Fe abundance is not dependent on the strengths of Fe
{\small I} lines, nor on the ionization state, respectively.  The
range in $\log g$ values is 0.6 to 1.4.  Using the true distance
modulus of (m-M)$_{\rm o} = 19.67$ for the Sextans dwarf spheroidal
\citep{mateo98}, the range in $M_{\rm V}$ of our objects is $-1.7$ and
$-2.2$, which corresponds to $1.0 < \log g <1.3$ if the $Y_{2}$
isochrones for extremely metal-poor stars are adopted \citep{kim02}.
Thus, the spectroscopic gravity from the Fe lines agrees well with the
gravity estimates from the color-magnitude diagram, and the remaining
scatter is most likely due to errors in the Fe line analysis.  The
{\vt} values in our sample range from 2.2 to 2.7~{\kms}, which are
typical of the values found for evolved red giants in the Galactic
halo (e.g., McWilliam et al. 1995).

\subsection{Na to Fe-peak elements}

To compare the chemical abundance results with
previous DART studies, in particular \citet{shetrone03}, analyses
using the same line lists would be ideal. However, the metallicity
of our objects is significantly lower than the stars previously
studied, therefore our analyses are limited to only a small number of 
lines in common and those have comparatively large oscillator strengths. 
The line list used in this work is given in Table~\ref{tab:ew} with 
equivalent widths for the Sextans stars and the comparison star. 
We found 26 Fe {\small I} lines in common with \citet{shetrone03}, 
among which the $\log gf$ values of 18 lines agree well;  the other 
8 lines show differences of 0.05--0.19~dex, but these are not systematic. 

Amongst the other species, the numbers of absorption lines commonly
used here and by \citet{shetrone03} include four lines of Fe {\small
  II}, three of Mg {\small I}, two of Na {\small I}, Ca {\small I}, Ti
{\small II} and Ba {\small II}, and one of Ti {\small I}, Cr {\small
  I}, and Ni {\small I}. The $\log gf$ values of three of the four Fe
{\small II} lines, which are originally from \citet{moity83}, differ
by 0.10--0.25~dex from those of \citet{shetrone03}, but again these
differences are not systematic.  Our $\log gf$ values for the two Mg
{\small I} lines at 5172 and 5183 {\AA}, originally from
\citet{aldenius07}, are 0.06--0.07~dex lower than in
\citet{shetrone03}. For the Ba {\small II} 5853~{\AA} line, our $\log
gf$ value is 0.10~dex higher than in \citet{shetrone03}.  These atomic
data differences will cause only small differences in the final
abundances between the two studies, smaller than the measurement
errors.  The $\log gf$ values for two of the Ti {\small II} lines in
this analysis are lower by about 0.25~dex compared to those in
\citet{shetrone03}.  The atomic data of most of the Ti {\small II}
lines are from \citet{ryabchikova94} and/or \citet{pickering01}, and
when lines are in common between these two studies, the atomic data
values agree very well.  We determined the Ti abundances using our
line list with no modification.

The iron abundances are determined from 42--61 Fe {\small I} and 2--7
Fe {\small II} lines. The number of lines used in the analysis is
dependent on the strengths of lines and the S/N ratio of the
spectrum. For instance, the numbers of Fe {\small II} lines used are
only two and three for S~15--19 (having the lowest metallicity) and
S~14--98 (having the lowest S/N ratio), respectively.
The derived iron abundances range from [Fe/H] = $-3.10$ to $-2.66$,
confirming that these stars are extremely metal-poor.  The iron abundance
of the comparison star HD~88609 is [Fe/H]$=-2.92$, which is in good agreement
with previously published values \citep{fulbright04,honda07}.

\begin{figure*}
   \centering
   \includegraphics{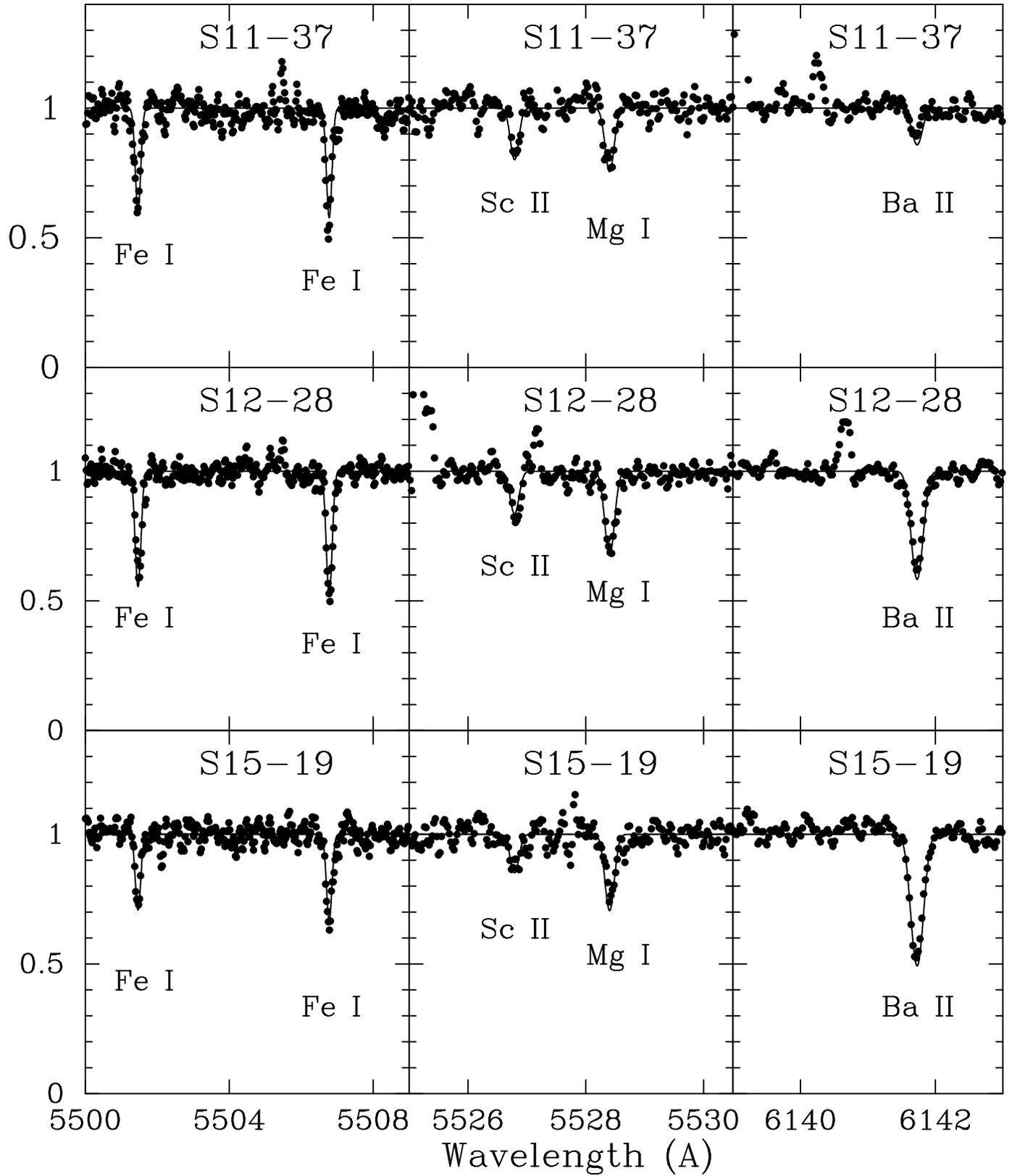}
   \caption{Observed spectra of the three Sextans objects S~11-37,
S12-28, and S15-19 for three wavelength ranges (dots). Synthetic
spectra calculated for the derived Mg, Sc, Fe and Ba abundances
(Table~\ref{tab:abund}) are also shown.  }
              \label{fig:sp}%
    \end{figure*}

\begin{figure}
   \centering
   \includegraphics[width=8.5cm]{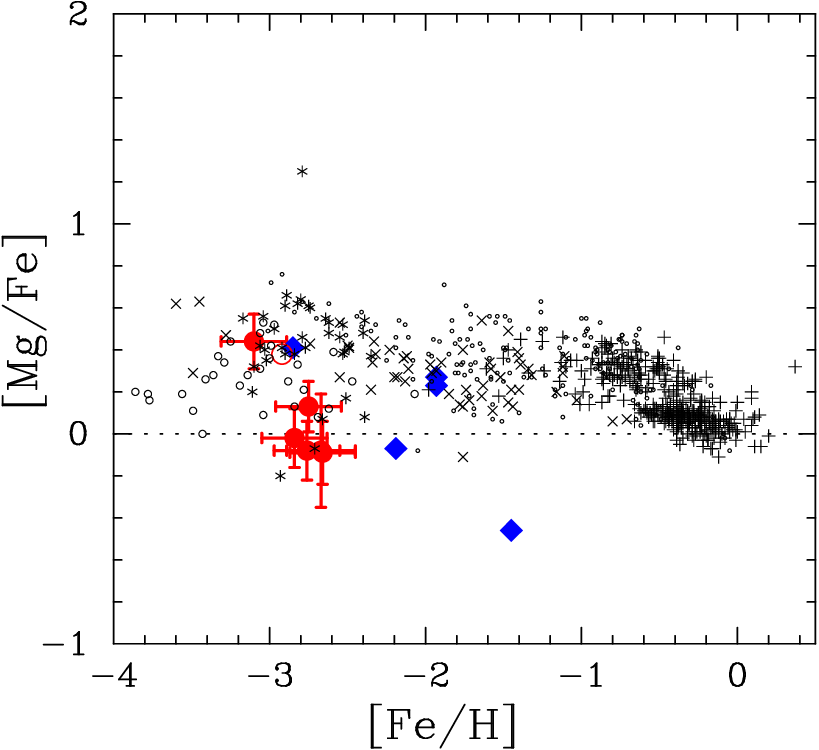}

      \caption{Mg abundance ratio ([Mg/Fe]) as a function of iron
  abundance ([Fe/H]). Our Sextans stars are depicted by filled
  circles with error bars, while other Sextans stars studied by
  \citet{shetrone01} are shown by filled diamonds. Our result for the
  comparison star HD~88609 is plotted by the large open circle. The
  abundances of Galactic halo stars are shown by open circles
  \citep{cayrel04}, asterisks \citep{honda04, aoki05}, crosses
  \citep{stephens02}, and small circles \citep{fulbright00}, while thick
  and thin disk stars studied by \citet{reddy03, reddy06} are shown by
  plus symbols.}

         \label{fig:mg}
   \end{figure}

The Mg abundances are determined from 3--4 of the four Mg {\small I} lines 
($\lambda$4571, 5172, 5183, and 5528).  Examples of the Mg {\small I} $\lambda$5528 
absorption lines are shown in Figure~\ref{fig:sp}, along with 
examples of Fe {\small I}, Sc {\small II}, and Ba {\small II} lines.  
The agreement in the abundances derived from the individual Mg {\small I} 
lines in fairly good agreement for each object: the standard deviation 
in the Mg abundances is less than 0.1~dex in most cases. 
We note that only two Mg {\small I}
lines at 5172 and 5183~{\AA} were used in the analysis of S~14--98
because of the low S/N ratio of the spectrum.  
The Mg/Fe ratios are shown in Figure~\ref{fig:mg}.  The Mg/Fe ratios 
for 5 out of 6 of our objects are, surprisingly, as low as the solar value,
i.e., lower than the typical values found in the Galactic halo stars by
0.3--0.4~dex.\footnote{The Mg/Fe ratios of extremely metal-poor red giants
in the Galactic halo are shown in Figure~\ref{fig:mg}.  These values are 
adopted from \citet{cayrel04}, \citet{honda04} and \citet{aoki05}. 
As discussed by \citet{aoki05}, there is a small systematic differences 
in the [Mg/Fe] values between \citet{cayrel04} and the other two papers:
over the metallicity range of $-3.2<$[Fe/H]$<-2.5$, which is that studied
here, the average of the [Mg/Fe] = +0.30 in \citet{cayrel04}, while
it is +0.42 in \citet{honda04} and \citet{aoki05}.   We note that
the remarkably Mg-enhanced ([Mg/Fe]$=$+1.25) star BS~16934--005 
\citep{aoki07b} is excluded from the sample of \citet{aoki05}.}  
The exception is S~15--19, with [Mg/Fe]=0.4. 
The [Mg/Fe] ratio of the comparison star HD~88609 is in good agreement
with previous studies and with the typical metal-poor Galactic halo
star ($=0.4$).

The difference in the Mg/Fe ratio between S~15--19 and other stars 
in Sextans is confirmed by comparing the spectral features in Figure~\ref{fig:sp}.
Comparisons are made for S~11--37, S~12--28, and S~15--19, which have
similar atmospheric parameters ({\teff}$\sim 4600$~K and {\logg}$\sim
1.3$). Synthetic spectra calculated for the derived Mg, Sc, Fe and Ba
abundances (Table~\ref{tab:abund}) are also shown in the figure. The
strength of the Mg {\small I} absorption lines at 5528~{\AA} are similar
between S~12-28 and S~15-19, while that of S~11--37 is slightly weaker.
In contrast, the Fe {\small I} lines at 5501 and 5507~{\AA} of S~15--19 
are weaker than those of the other stars.  As a result, the Fe abundance 
of S~15--19 is the lowest, and its Mg/Fe ratio is the highest.
The Mg/Fe ratios are discussed in detail in \S~\ref{sec:disc}.

\begin{figure}
   \centering
   \includegraphics[width=8.5cm]{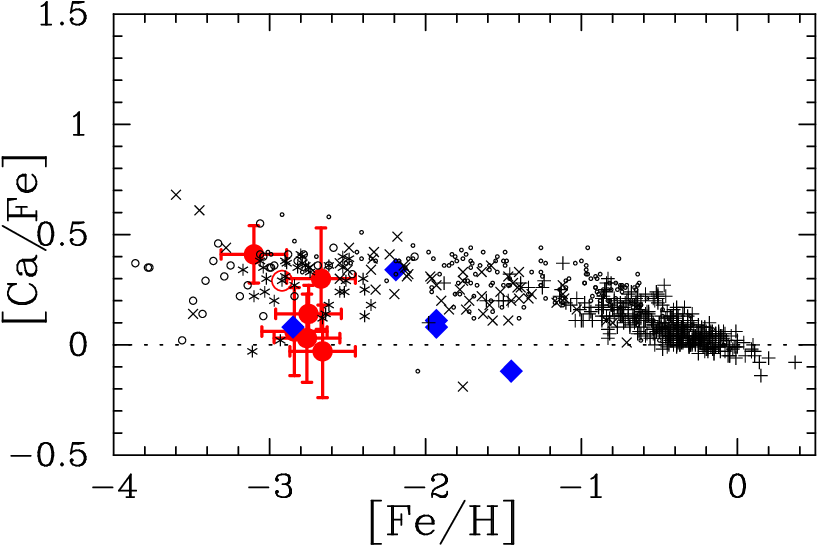}
   \includegraphics[width=8.5cm]{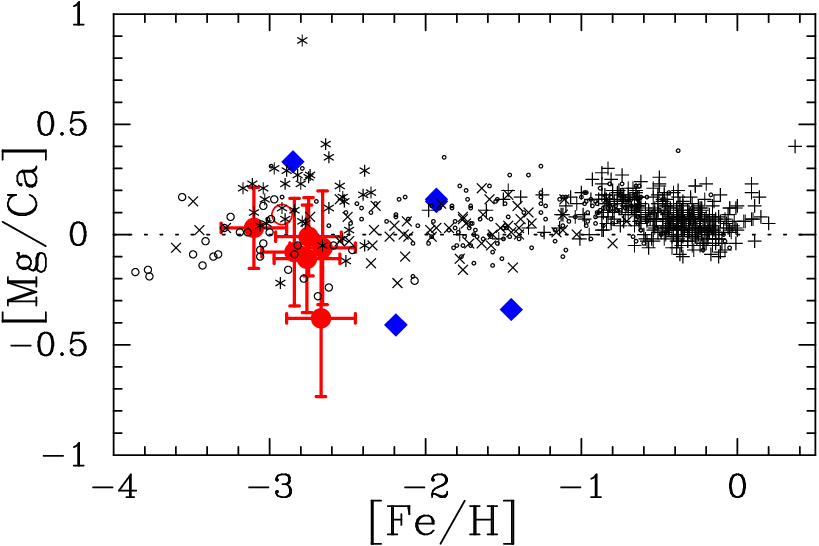}
      \caption{The same as Fig.\ref{fig:mg}, but for [Ca/Fe] (top) and [Mg/Ca] (bottom).
              }
         \label{fig:ca}
   \end{figure}

The Ca abundances are derived from 2--5 Ca {\small I} lines, including
the $\lambda$6122 and 6162 features that are detected in all of our
Sextans stars. The [Ca/Fe] and [Mg/Ca] ratios are shown in
Figure~\ref{fig:ca}.  Low Ca/Fe ratios are found for objects with 
low Mg/Fe values.   The only exception is S14-98, which shows a high Ca/Fe 
ratio ([Ca/Fe]=+0.27), but from the lowest S/N ratio spectrum, thus it
has the largest abundance uncertainty.  
The [Mg/Ca] values of our sample are solar ($\sim 0.0$), again with
S14-98 as the only exception probably due to its larger observational
errors (Fig.~\ref{fig:ca}).

The Na abundances are determined from the Na {\small I} D lines. The
strength of these lines are similar in most objects, and the derived
Na/Fe ratios agree with the solar value within the measurement
errors. The exception is S~10--14, which shows weaker absorption
lines.  A large non-LTE effect on the D line absorption is predicted 
for stars in this atmospheric parameter range. The non-LTE
corrections to the Na abundance derived from the D lines 
with $W=150$~m{\AA} in red giants is $\sim -0.5$ dex
\citep{takeda03, andrievsky07}. 
This correction is not included in our results. 
Since the effect is essentially systematic, this cannot be the
source of the difference of the Na abundances in S~10--14 and 
that of the other stars in our Sextans sample.

\begin{figure}
   \centering
   \includegraphics[width=8.5cm]{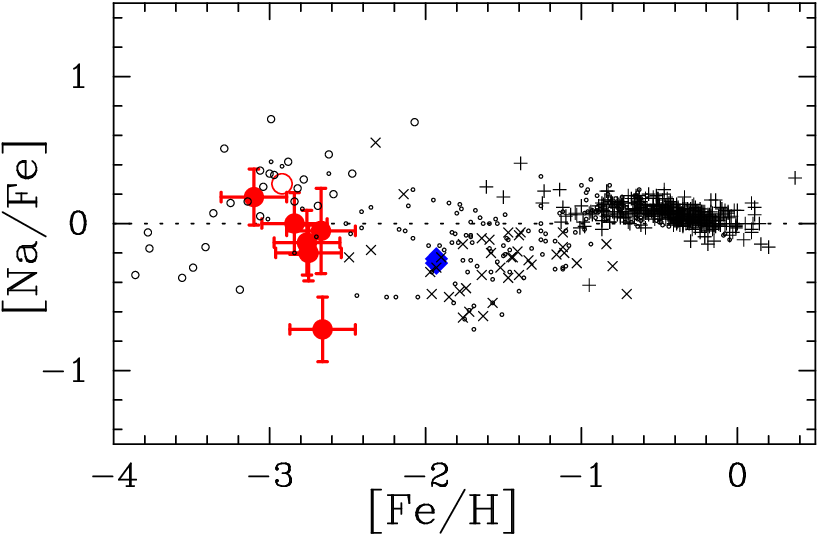}
   \includegraphics[width=8.5cm]{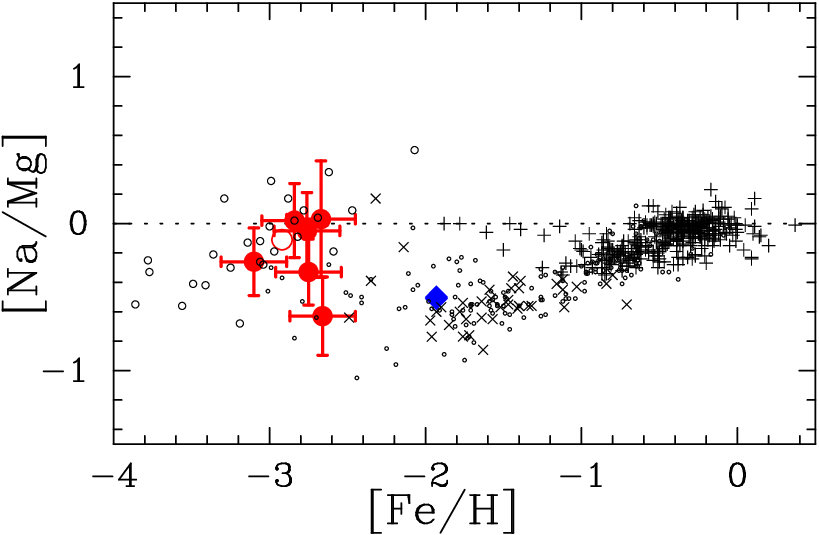}
      \caption{The same as Fig.\ref{fig:mg}, but for [Na/Fe] (top) and [Na/Mg] (bottom).}
         \label{fig:na}
   \end{figure}

Figure~\ref{fig:na} shows the [Na/Fe] and [Na/Mg] ratios for the
Sextans sample and Galactic stars.   Most stars have had their Na 
abundances determined from an LTE analysis of their D lines.
The Na abundances of the five objects in our sample have abundances
similar to Galactic halo stars ([Na/Fe]$\sim 0.0$).
The only exception is S~10--14, with an exceptionally low Na abundance. 
On the other hand, the plot of Na/Mg ratio shows a continuous
distribution from [Na/Fe]$=-0.6$ to $+0.1$.

Cr is under-abundant in all objects, as found for Galactic halo red
giants. The abundances are determined from the Cr {\small I} lines at
5206 and 5208~{\AA}. A dependence of the derived Cr abundance on the
stellar type (giant or dwarf) and on the species used in the analysis
(Cr {\small I} or {\small II}) was recently discussed by \citet{lai08}. 
Hence, we do not discuss the [Cr/Fe] values, other than to note that  
our Cr abundances are in good agreement with those of similar red giants
in the Galactic halo.

\subsection{The neutron-capture element Ba}

\begin{figure}
   \centering
   \includegraphics[width=8.5cm]{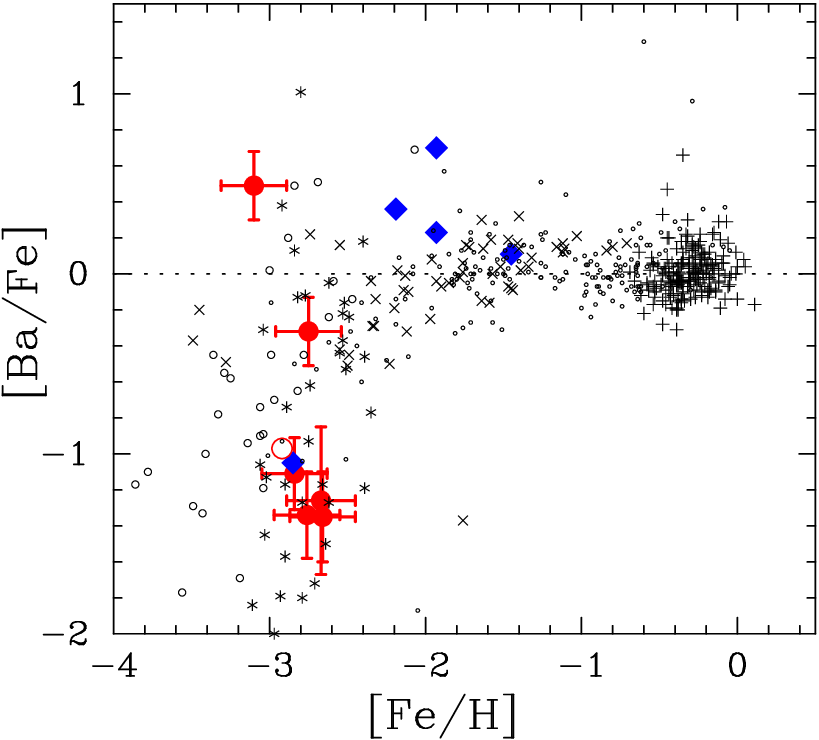}
      \caption{The same as Fig.\ref{fig:mg}, but for [Ba/Fe]. The open
circles mean the results of \citet{francois07} here.}
         \label{fig:ba}
   \end{figure}

The Ba abundance is determined from the three Ba {\small II} lines at
4934, 5853, and 6141~{\AA}. The resonance line at 4554~{\AA} is
not used because it is affected by a bad column on the CCD detector.
The effects of hyperfine splitting and isotope shifts are included in
the calculation, assuming the Ba isotopic ratio of the r-process
component of solar-system material. The Ba/Fe abundance ratios are
shown in Figure~\ref{fig:ba}.

Four of the Sextans stars are significantly under-abundant in
barium ([Ba/Fe]$\sim -1.3$), with abundance ratios in very good
agreement with each other.  In contrast, S~15--19 shows a large 
excess ([Ba/Fe]$=+0.5$) and S~12-28 shows a moderate excess 
([Ba/Fe]$=-0.3$), with respect to the other four stars. 
These large differences are clearly seen in the strength of the 
Ba {\small II} $\lambda$6141 alone, as shown in Figure~\ref{fig:sp}.

To constrain the origin of the Ba excess in S15-19, a measurement of
Eu abundance would be helpful, however no Eu line is detected in our
spectrum of this object. The upper limit estimated from the Eu {\small
  II} 6645~{\AA} line is [Eu/Fe]$<+1.6$.  This yields a lower limit of
[Ba/Eu]$>-1.1$, which does not exclude the s-process nor the r-process
as the origin of the Ba excess.

Another constraint might be the carbon abundance, because stars
showing large excesses of s-process elements usually show a carbon 
excess.   An upper limit to the carbon abundance in S~15--19 
can be estimated from the C$_{2}$ Swan 0--0 band at 
5165~{\AA}, as [C/Fe] $<+1.6$.  
Hence, this star is not extremely carbon-rich. 
However, taking into account the decrease in carbon expected
during the evolution of a red giant, as is found in Galactic 
halo stars (e.g. Spite et al. 2005, Aoki et al. 2007a), 
then this upper limit does not exclude the possibility of 
contamination in the s-process abundances through mass transfer 
from an AGB star. 

\subsection{Uncertainties}

Random errors in our abundance measurements from equivalent-widths are
estimated to be $\sigma N^{-1/2}$, where $\sigma$ is the standard
deviation in the abundances derived for individual lines and $N$ is the
number of lines used in the analysis. When the number of lines is
smaller than three, the standard deviation in the Fe {\small I} 
lines ($\sigma_{\rm Fe}$) is adopted. 
The random errors are less than 0.1~dex
for the iron abundance from Fe {\small I} lines, while typical values
for other species are 0.1--0.2~dex. The random errors for S~14--98 are
slightly larger due to the low S/N ratio of its spectrum.

To estimate the abundance errors due to the uncertainties in the
atmospheric parameters, we examine the comparison star HD~88609,
as shown in Table~\ref{tab:err}.  The results show the abundance 
changes in $\log \epsilon$ (Fe) for Fe {\small I} and {\small II}, 
and those in [X/Fe] for other species.  Total
uncertainties are obtained by adding these values, in quadrature, to
the random errors, and are listed in Table~\ref{tab:abund}.  The 
dominant abundance errors are due to the uncertainty in the 
effective temperatures in the total error of [Fe/H], while random 
errors are important for [X/Fe], in particular when the number of 
lines used in the analysis is small.

\section{Discussion and concluding remarks}\label{sec:disc}

\subsection{Metallicity distribution}

Our analysis of the high resolution spectra of six extremely
metal-poor star candidates in the Sextans dwarf galaxy confirms 
that these stars are extremely metal-poor, with [Fe/H] 
between $-2.6$ and $-3.1$.
Previously, only three stars with such low metallicities,
as confirmed by high resolution spectral analyses, were known 
in dwarf spheroidal galaxies: 
Draco 119 \citep{shetrone01,fulbright04}, 
Ursa Minor COS~4 \citep{sadakane04}, 
and Sextans S49 \citep{shetrone01}.
Therefore, this analysis provides the chemical composition for the 
largest sample of extremely metal-poor stars in dwarf galaxies to date.

\begin{figure}
   \centering
   \includegraphics[width=8.5cm]{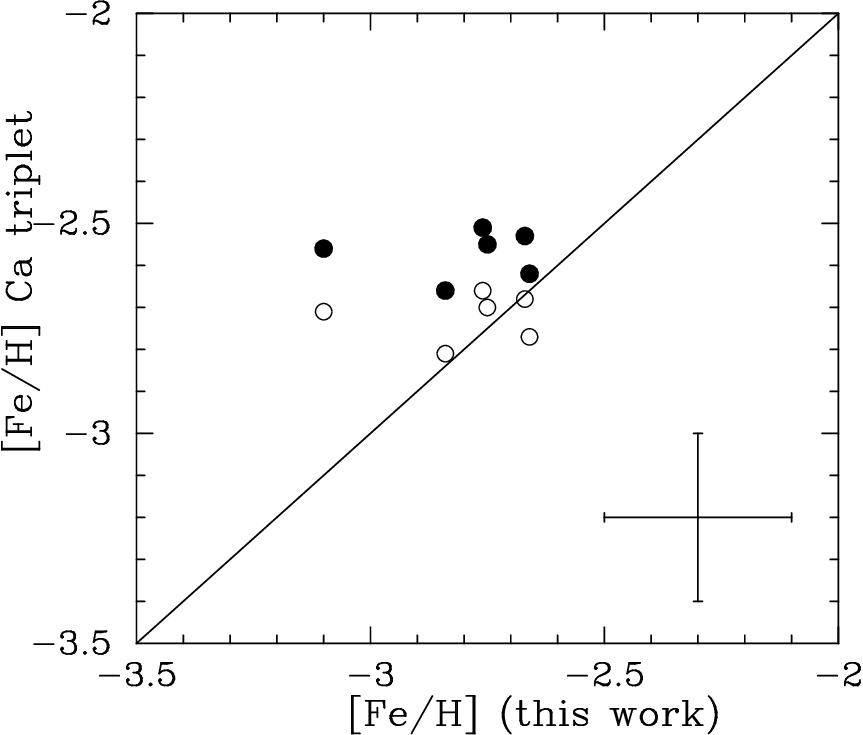}

      \caption{Comparison of [Fe/H] derived by this work and results
        from by the Ca triplet by \citet{helmi06} (filled circles) and
        \citet{battaglia08} (open circles). The typical errors are
        presented in the plot.  The agreement between the estimates
        from the Ca triplet and from the high resolution spectroscopy
        is good, with the exception of S~15--19 for which a lower
        [Fe/H] is derived from our high resolution study.}
         \label{fig:iron}
   \end{figure}

Our results have confirmed the low metallicities determined for these
stars from the VLT/FLAMES Ca triplet measurements
\citep{helmi06}.  However, Table~\ref{tab:iron} shows that the Fe
abundances that we derive are in fact slightly lower than those from
the Ca triplet estimates (also see Fig. \ref{fig:iron}).  The
agreement is fairly good for S10-14, while the difference is
significant for S15-19.  The average difference is 0.22~dex.  This
discrepancy is within the uncertainties in our {\teff} scale, however
it could also be an effect of the lack of calibrating clusters at
these metallicities of the Ca triplet method.

%, and therefore support the conclusion that stars 
%with metallicities below [Fe/H]$\sim -3$ are deficient in this galaxy.

The new calibration of metallicity from the Ca triplet lines by
\citet{battaglia08} (see \S~1) results in a better agreement of the Fe
abundance with that from our high resolution spectral analysis.
Table~\ref{tab:iron} also gives the metallicity estimated from the
calibration of \citet{battaglia08} for our sample.  These Fe values
are systematically lower than those of \citet{helmi06} by 0.15~dex, in
excellent agreement with our results.  The exception is S~15-19
([Fe/H]$=-3.1$ from the high resolution spectrum), for which a large
discrepancy still remains. The metallicity range of [Fe/H]$<-2.8$ is
not covered by the calibration of \citet{battaglia08} either.  High
resolution spectroscopy is still required to obtain reliable
metallicities for individual stars in this metallicity range at
present.  However, our results indicate that stars having
  [Fe/H]$<-3$ are indeed deficient in the Sextans dwarf spheroidal as
  discussed by \citet{helmi06}, though this should be confirmed by
  future studies for a larger sample.

Quite recently, \citet{kirby08} reported the metallicity distribution
for eight ultra-faint dwarf spheroidals around the Galaxy. The low
metallicity tail derived in their analysis, based on medium resolution
spectra of the Fe lines near the Ca triplet lines, is similar to that
of the Galactic halo, rather than to the classical dwarf spheroidals 
studied by \citet{helmi06}, including Sextans. 
 The existence of stars with [Fe/H] $\lesssim -3$ was
  confirmed from a high resolution spectral analyses by \citet{frebel09} 
  for stars in the two ultra-faint dwarf galaxies, Ursa Major II and Coma
  Berenices.
\citet{kirby08} have suggested that the difference in the low 
metallicity tail of the metallicity distribution functions between
the two classes of dwarf galaxies is real, and that the
ultra-faint dwarf spheroidals might be remnants of the building 
blocks of the Galactic halo.
It is likely that the classical dwarf galaxies, like Sextans, 
already are showing chemical enrichment due to its star formation
and chemical evolution history. 
Therefore, differences in the metallicity distributions, 
as well as detailed chemical composition, found for individual 
classical dwarf spheroidals could be the result of their own
independent evolutionary histories. 
For example, the classical dwarf galaxies may have had gas infall 
to explain the shortage of extremely metal-poor stars, similar to
one likely solution for the G-dwarf problem in the solar 
neighbourhood \citep{pagel75}.

 We emphasize again that, although the metallicity
  distributions of both ultra-faint and classical dwarf spheroidals
  have been explored by recent studies including the present work,
  further measurements of extremely metal-poor stars with high
  resolution spectroscopy are required to accurately determine the
  fraction of lowest metallicity ([Fe/H]$\lesssim -3$) stars.

\subsection{$\alpha$/Fe ratios}

The most surprising result of our study is the low Mg/Fe ratios
([Mg/Fe]$\sim 0.0$) found in five stars with respect to stars
in the Galactic halo. The Ca/Fe ratios are also low in these
objects, indicating that the abundance ratios between
$\alpha$-elements and iron in most of the extremely metal-poor stars 
in this dwarf galaxy are lower than those found in Galactic halo stars.
 The $\alpha$/Fe ratios of extremely metal-poor stars in the
  ultra-faint dwarf spheroidals studied by \citet{frebel09} are as
  high as those found in the bulk of Galactic halo stars. Therefore,
  the dwarf spheroidal Sextans does appear to have had a different 
  chemical evolution compared to the ultra-faint dwarfs, whether we
  examine abundance ratios or the metallicity distribution.

Low Mg/Fe ratios are found in some globular cluster stars.
Measurements of the Mg isotopic abundances from MgH absorption
features by \citet{shetrone96b} revealed that several red giants in
M13 have low $^{24}$Mg abundances, resulting in low Mg/Fe ratios.
However, globular cluster stars with low Mg also show overabundances
of Al and Na (e.g. Shetrone 1996a; see also Kraft 1994 for more on the
Na-O anti-correlation).  Unfortunately, Al abundances are not
available in our Sextans sample, and measurements of the Mg isotope
ratios are almost impossible because of the weakness of MgH features.
Future measurements of Al abundances from blue spectra will be useful
to constrain the possible effect of the MgAl chain. However, since the
Sextans stars are not overabundant in Na,  we do not expect such
abundance anomalies in this dwarf galaxy.

Low $\alpha$/Fe ratios are also found in dwarf galaxy stars at higher
metallicity ([Fe/H]$>-2$). For instance, [$\alpha$/Fe] ratios of stars
in the Sculptor dwarf galaxy show a decreasing trend with increasing
metallicity ([Fe/H]$>-2$) and reach [$\alpha$/Fe]$\sim -0.3$ at
[Fe/H]$=-1$ \citep[e.g., ][]{venn08}.
Several scenarios have been proposed to explain such low
Mg/Fe ratios, or their decreasing trend with increasing metallicity,
in dwarf galaxy stars. One idea is the contribution of type Ia
supernovae, which provide a large amount of iron, to metal-poor stars in
dwarf galaxies. If a low star formation rate is assumed for dwarf
galaxies, with respect to the Galactic halo and thick disk, the effects
of Type Ia supernovae could appear at lower metallicity. A similar
result is expected if one assumes that the material enriched by
supernova ejecta is lost from dwarf galaxies by some mechanism
(e.g. interaction with the Galaxy).

The low Mg/Fe ratios in extremely metal-poor stars found in the 
present study are not likely to be explained by contributions from 
Type Ia supernovae. 
The chemical abundance pattern of stars with such very low metallicity 
is expected to represent one or only a few nucleosynthesis events. 
Given the long time-scale of the Type Ia supernovae estimated from 
chemical evolution models ($>0.5$ billion years: e.g., Yoshii et al. 1996) 
and the low Type Ia supernova rate at low metallicity predicted by model
calculations \citep{kobayashi98}, then large contributions of this type 
of supernovae to extremely metal-poor stars in Sextans are not expected
(even though the progenitors and nature of Type Ia supernovae are still 
controversial). This is supported by the low Ba abundances in the low
Mg/Fe stars, indicating no significant contributions from 
intermediate-mass AGB stars that yield neutron-capture elements
via the s-process. The time-scale of enrichment from intermediate 
mass stars is similar to, or shorter than, that from Type Ia supernovae. 
Hence, our conclusion is that the low Mg/Fe abundances of five out of 
the six stars studied here are not a result of contributions from
Type Ia supernovae.

The low Mg/Fe ratios could be due to Type II supernovae instead, those
that yield a smaller amount of $\alpha$-elements in this galaxy.  For
example, model calculations of massive star evolution and supernova
nucleosynthesis predict that less massive stars
(e.g. $\sim$10~M$_{\odot}$) produce lower Mg/Fe ratios \citep[e.g.,
][]{woosley95}.  Either the upper IMF is truncated, or perhaps the
Type II supernovae ejecta from the highest masses is lost from the
galaxy at early times.  It should be noted that the existence of stars
with high Mg/Fe (S15-19 in our sample, and S49 in Shetrone et
al. 2001) suggests some range in the masses of Type II supernovae
progenitors in this galaxy.

Whatever processes are responsible for the low $\alpha$/Fe ratios in
the majority of our Sextans stars, these abundance ratios are clearly 
different from the majority of Galactic halo stars of the same metallicity. 
This result implies that classical dwarf spheroidal galaxies like Sextans 
do not resemble the building blocks of the main part of the Galactic halo, 
supporting the conclusion derived from the metallicity distribution 
functions.  It should be noted, however, that a small number of extremely 
metal-poor stars in the Galactic halo also show low $\alpha$/Fe ratios 
(e.g. McWilliam et al. 1995). These stars might have their origins in
classical dwarf spheroidals that merged during some later phase of the
Galactic halo formation.  Further searches for low $\alpha$/Fe stars in 
the Galactic halo and detailed studies of their kinematics would help
to address this question.

\subsection{Enrichment of neutron-capture elements}

The Ba is under-abundant in five extremely metal-poor stars in Sextans,
including S49 that was studied by \citet{shetrone01}. The Ba/Fe ratios
of these stars are similar to that of HD~88609, which represents
stars in the Galactic halo that are not strongly enriched by the
r-process. In contrast, the remaining two extremely metal-poor stars
in our sample show large and moderate excesses of this element. 

Measurements of heavy neutron-capture elements for Galactic halo stars
have revealed a rapid increase around at [Fe/H]$\sim -3$
\citep{mcwilliam95, honda04, aoki05, francois07}. 
The reason for this increase, in a very narrow range of metallicity, 
is unclear, but it does provide a strong constraint on the astrophysical 
site(s) of the r-process (e.g. Truran et al. 2002). 
One interpretation for the Ba enrichment at this metallicity is 
that the progenitors of core-collapse supernovae in which the 
r-process takes place are less massive stars ($\sim 10 $M$_{\odot}$), 
thus they only start to contribute at [Fe/H]$=-3$.  
It is interesting that a similar enrichment pattern of Ba is found in 
the Sextans sample.  
The origin of the Ba excesses in the two Ba-enhanced stars 
could then be due to the r-process.   If so, then this could have 
an impact on our understanding of the r-process, in general, since
the star formation histories are expected to have been quite different
between the Galactic halo and the classical dwarf galaxies. 
For example, perhaps the progenitor mass is not the critical
condition for the r-process, but instead the metallicity is the key 
to the r-process events. To confirm this hypothesis, measurements of 
other neutron-capture elements for the Ba-enhanced star S15-19 would
be extremely helpful.

The high Ba abundances in the three Sextans stars at [Fe/H]$\sim -2$
\citep{shetrone01} might be explained by a large contribution of the 
r-process with respect to Fe enrichment by Type II and/or Type Ia 
supernovae.  However, the Ba/Nd ratios measured by \citet{shetrone01} 
([Ba/Nd]$=-0.13$ to $+0.01$) are between the values expected from the 
r-process ($\sim -0.35$) and the s-process ($\sim +0.15$). 
Thus, it is necessary to estimate the s-process versus the r-process
contributions, e.g., through measurements of Eu for these objects.

\subsection{Summary and concluding remarks}

Our measurements of chemical abundances for six extremely metal-poor
red giants in the Sextans dwarf spheroidal galaxy have revealed that
the metallicity distribution of the low metallicity end and
$\alpha$/Fe abundance ratios are significantly different from those of
the main part of the Galactic halo. These results provide a new
constraint on the understanding of the formation and evolution of this
galaxy as well as on the roles of dwarf galaxies in the early Galaxy
formation. In contrast, the Ba abundance ratios in these stars show
similar scatter to that found in field stars. To understand
consistently these difference and similarity between this dwarf galaxy
and the Galactic halo, further observational studies for
Sextans and other dwarf galaxies are strongly desired.

\begin{acknowledgements}
This work has been supported in part by a Grant-in-Aid for
Scientific Research by the Japanese Ministry of Education, Culture,
Sports, Science and Technology (No. 19540245). W.~A. is supported by a
Grant-in-Aid for Science Research from JSPS (grant 18104003). AH
thanks the Netherlands Foundation of Scientific Research (NWO) for
financial support.  KAV acknowledges funding from the National 
Sciences and Engineering Research Council of Canada.
\end{acknowledgements}

\clearpage

%\end{document}

%                                     Two column figure (place early!)
%______________________________________________ Gamma_1 (lg rho, lg e)

\clearpage

%
%                                                One column figure
%----------------------------------------------------------- S_vib
%
%                                                One column figure
%----------------------------------------------------------- S_vib
%
%______________________________________________________________
%                                                One column figure
%----------------------------------------------------------- S_vib
%______________________________________________________________
%                                                One column figure
%----------------------------------------------------------- S_vib
%
%______________________________________________________________
%                                                One column figure
%----------------------------------------------------------- S_vib
%
%_____________________________________________________________
%                                             Two column Table 
%_____________________________________________________________
%

%\clearpage

\begin{table*}
\caption{Objects and observation details}             
\label{tab:obj}      
\centering          
\begin{tabular}{llllrrll}
\hline\hline       
Star & RA(2000) & Dec(2000) & Observation & exposure$^{\mathrm{a}}$  & S/N$^{\mathrm{b}}$  & $V_{\rm H}$({\kms}) & HJD  \\

\hline                    
S~10--14 & 10 13 34.70 & $-$02 07 57.9 & Jan 2007 & 330 (11) & 36 & $233.86 \pm 0.08$ & 2454126 \\
S~11--13 & 10 11 42.96 & $-$02 03 50.4 & Jan 2007 & 60  (2)  & 21 & $224.71 \pm 0.08$ & 2454128 \\
S~11--37 & 10 13 45.48 & $-$01 56 16.3 & Jan 2007 & 240  (8) & 32 & $221.79 \pm 0.08$ & 2454128 \\
S~12--28 & 10 11 17.15 & $-$02 00 24.0 & Jan 2007 & 240  (8) & 39 & $201.96 \pm 0.05$ & 2454126 \\
S~14--98 & 10 13 24.48 & $-$02 12 03.5 & Jan 2007 & 85   (3) & 17 & $220.7 \pm 0.13$ & 2454128 \\
S~15--19 & 10 11 26.92 & $-$02 05 41.7 & May 2005 & 390 (13) & 46 & $226.05 \pm 0.11$ & 2453511 \\
%S~24--72 & 10 15 03    & $-$01 29 55 & Jan 2007 & 180    (6) & 34 & $210.04 \pm 0.04$ & 2454129.05 \\
HD~88609 & 10 14 28.98 & $+$53 33 39.4 & Jan 2007 & 10.5 (4) & 440 & $-37.58 \pm 0.02$ & 2454126 \\
\hline
S~10--12$^{\mathrm{c}}$ & 10 13 52.1  & $-$02 02 55.1 & Jan 2007 & 30 (1) & 11 & ... & 2454126 \\
S~11--36$^{\mathrm{c}}$ & 10 13 45.5  & $-$01 56 16.3 & Jan 2007 & 240 (8) & 8 & ... & 2454126 \\
\hline                  
\end{tabular}
\begin{list}{}{}
\item[$^{\mathrm{a}}$] Total exposure time (minutes) and the number of exposures.
\item[$^{\mathrm{b}}$] S/N ratio per 1.8~{\kms} pixel at 5180~{\AA}.
\item[$^{\mathrm{c}}$] These stars were observed, but are not included in the present work (see text).
\end{list}
\end{table*}
%_____________________________________________________________
%
\begin{table*}
\caption{Photometry data and stellar parameters}             
\label{tab:param}      
\centering          
\begin{tabular}{lcccccccccc}
\hline\hline       
Star   & $V$   & $V-I$ & $K$ & {\ebv} & {\teff}$(V-K)$ & {\teff}$(V-I)$ & {\teff}(adopted) & {\logg} & [Fe/H] & {\vt} \\
       &       &       &     &        &  (K)           & (K)            & (K)              &         &        & {\kms} \\
\hline
S~10--14 & 17.64 & 1.08 & 15.08 & 0.038 & 4619 & 4700 & 4620 & 1.2 & $-$2.7 & 2.2 \\
S~11--13 & 17.53 & 1.15 & 14.71 & 0.038 & 4399 & 4559 & 4400 & 0.6 & $-$2.8 & 2.4 \\
S~11--37 & 17.96 & 1.06 & 15.34 & 0.038 & 4564 & 4742 & 4560 & 1.3 & $-$2.9 & 2.2 \\
S~12--28 & 17.52 & 1.08 & 14.91 & 0.038 & 4574 & 4700 & 4570 & 1.4 & $-$2.8 & 2.7 \\
S~14--98 & 18.06 & 1.00 & 15.61 & 0.038 & 4725 & 4877 & 4730 & 1.1 & $-$2.7 & 2.8 \\
S~15--19 & 17.54 & 1.01 & 14.98 & 0.038 & 4619 & 4854 & 4620 & 1.2 & $-$3.1 & 2.6 \\
%S24-72 & 17.42 & 1.24 & 14.49 & 0.038 & 4317 & 4397 & 4320 & 0.8 & $-$3.0 & 2.6 \\
HD~88609 & 8.58 & 1.12 &  6.01 & 0.000 & 4522 & 4517 & 4520 & 1.1 & $-$3.0 & 2.5 \\
\hline
\end{tabular}
\end{table*}

\addtocounter{table}{1}

\begin{table*}
\caption{Elemental abundances}             
\label{tab:abund}      
\centering          
\begin{tabular}{llccccccccccc}
\hline\hline       
Star     &  element & Fe   & Fe  & Na  & Mg  & Ca  & Sc  & Ti  & Ti  & Cr   & Ni  & Ba \\
         &  species & Fe {\small I} & Fe {\small II}  & Na {\small I}  & Mg {\small I}   & Ca {\small I}   & Sc {\small II}  & Ti {\small I}  & Ti {\small II}  & Cr {\small I}   & Ni {\small I}  & Ba {\small II} \\
\hline
Sun      & $\log \epsilon$ & 7.45  & 7.45  & 6.17  & 7.53  & 6.31  & 3.05  & 4.90 & 4.90  & 5.64  & 6.23 & 2.17 \\
\hline
S~10--14   & $\log \epsilon$ &  4.79 &  4.81 &  2.79 &  4.77 &  3.62 & ... & 2.43 &  2.12 &  2.37 & ... & -1.84  \\
         & N      & 58    &     5 &     2 &     4 &     2 & ... &     1 &    6 &    2  & ... &     2  \\
         & [X/Fe]$^{\mathrm{a}}$ & -2.66 & -2.64 & -0.72 & -0.09 & -0.03 & ... & 0.19 & -0.11 & -0.61 & ... & -1.35 \\
         & err    &  0.21 &  0.22 &  0.22 &  0.15 &  0.21 & ... & 0.28 &  0.19 &  0.20 & ... & 0.25 \\
\hline
S~11--13   & $\log \epsilon$ &  4.69 &  4.68 &  3.28 &  4.68 & 3.58 &  0.14 &  1.97 &  2.11 &  2.22 & 3.52 & -1.93 \\
         & N      &    55 &     5 &     2 &     4 &    2 &     1 &     2 &     7 &     2 &    1 &     2 \\
         & [X/Fe]$^{\mathrm{a}}$ & -2.76 & -2.77 & -0.13 & -0.08 & 0.03 & -0.14 & -0.17 & -0.03 & -0.66 & 0.05 & -1.34 \\
         & err    &  0.21 &  0.18 &  0.22 &  0.14 & 0.20 &  0.31 &  0.19 &  0.19 &  0.19 & 0.27 &  0.24 \\
\hline
S~11--37   & $\log \epsilon$ &  4.61 &  4.57 &  3.32 &  4.67 &  3.53 &  0.41 &  1.80 & 2.50 &  2.21 & 3.58 & -1.79 \\
         & N      &    50 &     6 &     2 &     4 &     2 &     1 &     1 &    4 &    2  &    2 &     2 \\
         & [X/Fe]$^{\mathrm{a}}$ & -2.84 & -2.88 &  0.00 & -0.02 &  0.06 &  0.20 & -0.26 & 0.44 & -0.59 & 0.20 & -1.11 \\
         & err    &  0.21 &  0.23 &  0.21 &  0.14 &  0.20 &  0.31 &  0.26 & 0.20 &  0.18 & 0.20 &  0.23 \\
\hline
S~12--28   & $\log \epsilon$ &  4.70 &  4.68 &  3.22 & 4.90 & 3.70 & 0.30 &  2.10 &  2.12 &  2.43 & 3.57 & -0.90 \\
         & N      &    61 &     7 &     2 &    4 &    4 &    3 &    3  &    13 &     1 &    1 &     3 \\
         & [X/Fe]$^{\mathrm{a}}$ & -2.75 & -2.77 & -0.20 & 0.13 & 0.14 & 0.01 & -0.05 & -0.03 & -0.46 & 0.10 & -0.32 \\
         & err    &  0.21 &  0.13 &  0.19 & 0.12 & 0.13 & 0.21 &  0.13 &  0.17 &  0.22 & 0.23 &  0.19 \\
\hline
%S14-98   & $\log \epsilon$ &  4.83 &  4.90 &  3.54 &  4.84 & 3.96 &  0.28 &  2.56 &  2.51 &  2.45 & 3.91 & -1.71 \\
S~14--98   & $\log \epsilon$ &  4.78 &  4.82 &  3.45 &  4.78 & 3.94 &  0.24 &  2.54 &  2.43 &  2.40 & ... & -1.76 \\
         & N      &    42 &     3 &     2 &     2 &    3 &     1 &     2 &     5 &     2 & ... &     1 \\
         & [X/Fe]$^{\mathrm{a}}$ & -2.67 & -2.63 & -0.05 & -0.08 & 0.30 & -0.17 &  0.31 &  0.20 & -0.57 & ... & -1.26 \\
         & err    &  0.22 &  0.25 &  0.29 &  0.27 & 0.23 &  0.42 &  0.27 &  0.23 &  0.27 & ... &  0.41 \\
%\hline
%S24-72   & $\log \epsilon$ &  4.50 &  4.47 & 4.23 & 4.59 & 3.55 & 0.27 &  1.93 &  1.94 &  2.32 & 3.41 & -2.08 \\
%4320     & N      &    65 &     6 &    3 &    5 &    5 &    3 &     4 &    12 &     2 &    2 &     2 \\
%         & [X/Fe] & -2.95 & -2.98 & 1.01 & 0.01 & 0.19 & 0.16 & -0.02 & -0.02 & -0.37 & 0.13 & -1.30 \\
%         & err    &  0.21 &  0.17 & 0.17 & 0.11 & 0.13 & 0.21 &  0.12 &  0.17 &  0.16 & 0.17 &  0.21 \\
\hline
S~15--19   & $\log \epsilon$ &  4.35 &  4.35 & 3.25 & 4.86 & 3.62 & 0.23 & 2.03 &  1.40 &  1.98 & 3.41 & -0.44 \\
         & N      &    47 &     2 &    2 &    3 &    5 &    1 &    1 &     1 &     2 &    2 &     3 \\
         & [X/Fe]$^{\mathrm{a}}$ & -3.10 & -3.10 & 0.18 & 0.44 & 0.41 & 0.28 & 0.23 & -0.40 & -0.56 & 0.28 &  0.49 \\
         & err    &  0.21 &  0.18 & 0.19 & 0.13 & 0.13 & 0.27 & 0.22 &  0.27 &  0.15 & 0.17 &  0.19 \\
\hline
HD~88609  & $\log \epsilon$ &  4.53 &  4.52 & 3.51 & 4.98 & 3.68 & 0.15 & 2.09 & 2.13 &  2.31 & 3.23 & -1.72 \\
         & N      &    65 &     8 &     2 &    5 &    5 &    3 &    6 &   14 &    2 &     2 &    4 \\
         & [X/Fe]$^{\mathrm{a}}$ & -2.92 & -2.94 & 0.27 & 0.38 & 0.29 & 0.03 & 0.11 & 0.15 & -0.40 & -0.08 & -0.97 \\
         & err    &  0.21 &  0.12 & 0.16 & 0.08 & 0.11 & 0.19 & 0.08 & 0.16 &  0.11 & 0.13 & 0.17 \\
\hline
\end{tabular}
\begin{list}{}{}
\item{$^{\mathrm{a}}$[Fe/H] values are given for Fe in these lines.}
\end{list}
\end{table*}

\clearpage

%_____________________________________________________________
%
\begin{table}
\caption[]{Abundance changes by changing stellar parameters}             
\label{tab:err}      
\centering          
\begin{tabular}{lcccc}
\hline\hline       
     & $\delta$({\teff}) & $\delta$({\logg}) & $\delta$([Fe/H]) & $\delta$({\vt}) \\
     & +150~K & +0.3~dex & +0.3~dex & +0.3~{\kms} \\ 
\hline
Fe {\small I}  & 0.19     & $-$0.04 & 0.00    & $-$0.06 \\
Fe {\small II} & 0.02     & 0.08    & 0.00    & $-$0.05 \\
Na {\small I} & 0.03     & $-$0.05 & $-$0.02 & $-$0.09 \\
Mg {\small I}  & $-$0.02  & $-$0.04 & 0.00    & $-$0.01 \\
Ca {\small I}  & $-$0.07  & 0.01    & 0.01    & 0.04 \\
Sc {\small II} & $-$0.12  & 0.12    & 0.00    & 0.04 \\
Ti {\small I}  & 0.00     & 0.00    & 0.01    & 0.04 \\
Ti {\small II} & $-$0.12  & 0.11    & 0.00    & $-$0.01 \\
Cr {\small I}  & 0.01     & $-$0.01 & 0.00    & $-$0.03 \\
Ni {\small I}  & $-$0.05  & 0.02    & 0.01    & 0.06  \\
Ba {\small II} & $-$0.08  & 0.12    & 0.00    & 0.05  \\
\hline
\end{tabular}
\end{table}

%_____________________________________________________________
%
\begin{table}
\caption[]{Abundance changes by changing stellar parameters}             
\label{tab:iron}      
\centering          
\begin{tabular}{lccc}
\hline\hline       
Object     & [Fe/H]      & \multicolumn{2}{c}{[Fe/H](Ca triplet)} \\
\cline{3-4}
           & (this work) & Helmi et al. (2006) & Battaglia et al. (2008)$^{\mathrm{a}}$ \\
\hline
S~10--14 & $-2.66\pm 0.21$ & $-2.62$ & $-2.77$ \\
S~11--13 & $-2.76\pm 0.21$ & $-2.51$ & $-2.66$ \\
S~11--37 & $-2.84\pm 0.21$ & $-2.66$ & $-2.81$ \\
S~12--28 & $-2.75\pm 0.21$ & $-2.55$ & $-2.70$ \\
S~14--98 & $-2.67\pm 0.22$ & $-2.53$ & $-2.68$ \\
S~15--19 & $-3.10\pm 0.21$ & $-2.56$ & $-2.71$ \\
%S24-72 & -2.95 & -2.80 \\
\hline
\end{tabular}
\begin{list}{}{}
\item{$^{\mathrm{a}}$Values calculated from the calibration of Battaglia et al. (2008).}
\end{list}
\end{table}

%
%_____________________________________________________________
%                              Table longer than a single page  
%  In the preamble, use:              \usepackage{longtable}
%-------------------------------------------------------------
%          All long tables have to be placed at the end, after 
%                                        \end{thebibliography}
%
% In the text, at the place where the large table should appear
% add the command:
\addtocounter{table}{1}
% Tables counters will be well numbered.
%
% If table 2
\longtab{3}{
\begin{longtable}{lcccccccccc}
\caption{\label{tab:ew} Equivalent widths}\\
\hline\hline
       & $\lambda$({\AA}) & $\chi$(eV) & log(gf) & S 10-14 & S 11-13 & S 11-37 & S 12-28 & S 14-98 & S 15-19 &  HD088609    \\
\hline
\endfirsthead
\caption{continued.}\\
\hline\hline
       & $\lambda$({\AA}) & $\chi$(eV) & log(gf) & S 10-14 & S 11-13 & S 11-37 & S 12-28 & S 14-98 & S 15-19 & HD088609    \\
\hline
\endhead
\hline
\endfoot
  Na I & 5889.95 &  0.00 &  0.12 & 124.6 & 158.3 & 153.0 & 168.6 & 165.0 & 153.6 & 179.4 \\
  Na I & 5895.92 &  0.00 & -0.18 & 90.7  & 149.4 & 137.3 & 142.2 & 149.9 & 137.6 & 156.9\\
  Mg I & 4571.10 &  0.00 & -5.39 & 52.7  & 62.3  & 61.1  & 79.9  & ... & ... & 81.1\\
  Mg I & 4702.99 &  4.33 & -0.38 & ... & ... & ... & ... & ... & ... & 63.6\\
  Mg I & 5172.68 &  2.71 & -0.45 & 163.1 & 175.4 & 158.7 & 190.9 & 152.8 & 184.1 & 195.5\\
  Mg I & 5183.60 &  2.72 & -0.24 & 175.1 & 182.3 & 164.9 &  229.6 & 201.9 & 190.4 & 223.2\\
  Mg I & 5528.40 &  4.35 & -0.34 & 54.4 & 52.7 & 56.2 & 67.3 & 55.7 & 55.4 & 69.2\\
  Ca I & 5588.75 &  2.53 &  0.20 & ...& ...& ...& 51.8 & ...& 45.0 & 43.3 \\
  Ca I & 6102.72 &  1.88 & -0.79 & ...& ...& ...& ...& 55.1 & 27.1 & 31.2 \\
  Ca I & 6122.22 &  1.89 & -0.31 & 44.4 & 60.7 & 47.6 & 51.4 & 58.8 & 40.7 & 60.0 \\
  Ca I & 6162.17 &  1.90 & -0.09 & 66.8 & 65.9 & 60.2 & 79.1 & 75.8 & 73.8 & 73.9 \\
  Ca I & 6462.57 &  2.52 &  0.31 & ...& ...& ...& 52.2 & ...& 32.5 & 50.6 \\
  Ti I & 4981.73 &  0.85 &  0.56 & ...& ...& 37.1 & 57.4 & 81.0 & 47.8 & 52.9 \\
  Ti I & 4991.06 &  0.84 &  0.44 & ...& ...& ...& 45.8 & ...& ...& 49.2 \\
  Ti I & 5007.21 &  0.82 &  0.17 & 48.4 & 35.0 & ...& 41.9 & 44.2 & ...& 41.7 \\
  Ti I & 5192.97 &  0.02 & -0.95 & ...& 30.5 & ...& ...& ...& ...& 31.9 \\
  Cr I & 5206.04 &  0.94 &  0.02 & 62.6 & 81.4 & 73.5 & 92.8 & 78.9 & 56.5 & 76.5 \\
  Cr I & 5208.42 &  0.94 &  0.16 & 97.2 & 91.8 & 79.9 & 104.4 & 90.1 & 66.8 & 102.5 \\
  Mn I & 4823.50 &  2.32 &  0.14 & ...& ...& ...& 28.1 & ...& ...& 15.7 \\
  Fe I & 4871.32 &  2.87 & -0.36 & 64.7 & ...& 50.6 & 79.8 & 128.2 & 50.0 & 66.1 \\
  Fe I & 4872.14 &  2.88 & -0.57 & 60.8 & ...& ...& 73.1 & ...& 45.9 & 55.9 \\
  Fe I & 4890.75 &  2.88 & -0.39 & 67.8 & 80.1 & 74.8 & 95.2 & ...& 42.2 & ...\\
  Fe I & 4891.49 &  2.85 & -0.11 & 82.7 & ...& 60.8 & 95.0 & ...& ...& 80.9 \\
  Fe I & 4903.31 &  2.88 & -1.08 & ...& ...& ...& ...& ...& ...& 38.8 \\
  Fe I & 4918.99 &  2.85 & -0.34 & 72.6 & 88.3 & 76.0 & 82.0 & 82.2 & 61.3 & 69.8 \\
  Fe I & 4920.50 &  2.83 &  0.07 & 106.1 & 114.3 & 92.7 & 98.1 & 89.7 & 79.5 & 92.2 \\
  Fe I & 4938.81 &  2.88 & -1.08 & ...& ...& ...& 47.9 & ...& ...& 32.5 \\
  Fe I & 4939.69 &  0.86 & -3.25 & 56.8 & ...& ...& 78.1 & 78.5 & ...& 63.2 \\
  Fe I & 4957.60 &  2.81 &  0.23 & ...& 106.1 & ...& ...& 105.6 & 85.0 & 112.4 \\
  Fe I & 4994.13 &  0.92 & -3.08 & ...& ...& 72.5 & ...& ...& ...& 74.8 \\
  Fe I & 5006.12 &  2.83 & -0.64 & 48.8 & 52.7 & 65.6 & ...& 68.3 & 46.5 & 57.6 \\
  Fe I & 5012.07 &  0.86 & -2.64 & 84.5 & 112.2 & 104.3 & 132.2 & 105.2 & 86.9 & 105.1 \\
  Fe I & 5041.76 &  1.49 & -2.20 & 68.3 & 71.2 & 79.0 & 104.1 & 86.8 & 53.2 & 83.4 \\
  Fe I & 5049.82 &  2.28 & -1.42 & 72.6 & 69.1 & 66.9 & 73.5 & 75.8 & 41.4 & 62.6 \\
  Fe I & 5051.63 &  0.92 & -2.80 & 92.7 & 102.9 & 101.2 & 110.0 & 70.6 & ...& 93.9 \\
  Fe I & 5068.77 &  2.94 & -1.23 & ...& 48.4 & 46.8 & 28.0 & ...& ...& 29.0 \\
  Fe I & 5079.22 &  2.20 & -2.07 & 42.0 & ...& ...& 54.0 & ...& ...& 33.8 \\
  Fe I & 5079.74 &  0.99 & -3.22 & 74.1 & 97.6 & ...& 85.2 & ...& 41.8 & 59.5 \\
  Fe I & 5083.34 &  0.96 & -2.96 & 77.2 & 106.6 & 80.6 & 90.1 & ...& 61.1 & 79.0 \\
  Fe I & 5123.72 &  1.01 & -3.07 & 75.3 & 86.1 & ...& 77.9 & ...& 37.6 & 70.6 \\
  Fe I & 5125.12 &  4.22 & -0.14 & 25.6 & ...& ...& ...& ...& ...& ...\\
  Fe I & 5127.36 &  0.92 & -3.31 & ...& 74.3 & 71.2 & 79.4 & 75.8 & 42.8 & 62.4 \\
  Fe I & 5133.69 &  4.18 &  0.14 & 34.8 & ...& ...& ...& ...& ...& 23.0 \\
  Fe I & 5142.93 &  0.96 & -3.08 & 87.0 & 91.4 & 76.1 & 79.3 & 79.8 & 48.8 & 69.8 \\
  Fe I & 5150.84 &  0.99 & -3.07 & 68.2 & 64.0 & 46.3 & 75.0 & ...& 45.9 & ...\\
  Fe I & 5151.91 &  1.01 & -3.32 & 57.0 & 73.2 & 58.7 & 64.5 & ...& 41.3 & 52.3 \\
  Fe I & 5162.27 &  4.18 &  0.02 & ...& ...& ...& 30.1 & ...& ...& 19.0 \\
  Fe I & 5166.28 &  0.00 & -4.20 & 92.0 & 80.2 & 77.0 & 110.2 & 83.5 & 73.7 & 86.4 \\
  Fe I & 5171.60 &  1.49 & -1.79 & 94.5 & 121.2 & 90.0 & 106.6 & 89.0 & 91.3 & 104.3 \\
  Fe I & 5191.46 &  3.04 & -0.55 & 63.7 & 59.5 & 63.3 & 54.7 & 44.0 & ...& 47.9 \\
  Fe I & 5192.34 &  3.00 & -0.42 & 60.3 & 70.0 & 61.3 & 62.3 & ...& ...& 56.3 \\
  Fe I & 5194.94 &  1.56 & -2.09 & 84.2 & 95.7 & 73.1 & ...& 72.1 & 52.4 & 83.1 \\
  Fe I & 5198.71 &  2.22 & -2.13 & ...& ...& ...& 50.3 & ...& ...& 25.9 \\
  Fe I & 5202.34 &  2.18 & -1.84 & 79.2 & 86.0 & ...& 57.0 & 104.4 & ...& 47.0 \\
  Fe I & 5216.27 &  1.61 & -2.15 & 79.9 & 99.3 & 69.1 & 83.4 & 65.3 & 49.2 & 74.3 \\
  Fe I & 5225.53 &  0.11 & -4.79 & 47.0 & 67.0 & ...& 42.1 & 61.0 & ...& ...\\
  Fe I & 5232.94 &  2.94 & -0.06 & 84.9 & 88.2 & 96.1 & 85.5 & 60.7 & 58.0 & 81.8 \\
  Fe I & 5250.65 &  2.20 & -2.05 & 38.0 & 65.1 & ...& 47.1 & ...& 28.5 & 31.1 \\
  Fe I & 5266.56 &  3.00 & -0.39 & 76.0 & 81.8 & 76.4 & ...& ...& 32.7 & 60.6 \\
  Fe I & 5269.54 &  0.86 & -1.32 & 174.5 & 156.3 & 149.2 & 180.1 & ...& 148.4 & ...\\
  Fe I & 5270.36 &  1.61 & -1.51 & 113.7 & 129.1 & 109.2 & 141.6 & ...& 108.3 & 133.6 \\
  Fe I & 5281.79 &  3.04 & -1.02 & ...& 48.8 & ...& 50.4 & ...& ...& 34.0 \\
  Fe I & 5324.18 &  3.21 & -0.24 & 80.8 & 80.3 & 66.9 & 77.3 & ...& 53.7 & 60.4 \\
  Fe I & 5328.04 &  0.92 & -1.47 & 138.7 & 175.0 & 132.1 & 160.6 & 122.3 & 159.4 & 158.7 \\
  Fe I & 5328.53 &  1.56 & -1.85 & 109.0 & 124.8 & 101.6 & 98.7 & 100.7 & 78.8 & 100.2 \\
  Fe I & 5332.90 &  1.56 & -2.94 & 61.3 & 67.1 & 45.4 & 43.1 & 36.1 & ...& 35.5 \\
  Fe I & 5339.93 &  3.27 & -0.68 & ...& 49.3 & ...& 37.8 & 67.8 & 26.0 & ...\\
  Fe I & 5341.02 &  1.61 & -2.06 & 98.8 & 106.5 & 75.3 & 105.5 & 103.3 & 81.6 & 89.4 \\
  Fe I & 5371.49 &  0.96 & -1.64 & ...& 170.5 & 137.5 & 154.9 & 154.7 & 120.5 & 147.8 \\
  Fe I & 5397.13 &  0.92 & -1.99 & 138.6 & 140.0 & 99.3 & 164.3 & 127.6 & 125.3 & 134.8 \\
  Fe I & 5405.77 &  0.99 & -1.84 & 92.8 & 157.5 & ...& 156.0 & 120.0 & 117.4 & 137.2 \\
  Fe I & 5429.70 &  0.96 & -1.88 & 127.2 & 141.9 & 123.6 & 141.6 & 153.4 & 107.8 & 140.3 \\
  Fe I & 5434.52 &  1.01 & -2.12 & 98.8 & 131.6 & 97.7 & 134.6 & 133.6 & 103.8 & 123.5 \\ 
  Fe I & 5446.92 &  0.99 & -1.93 & 127.1 & 145.6 & 131.6 & ...& 129.1 & 114.5 & ...\\
  Fe I & 5455.61 &  1.01 & -2.09 & 145.9 & 160.8 & 134.2 & 150.6 & 154.3 & 122.2 & 138.4 \\
  Fe I & 5497.52 &  1.01 & -2.85 & 88.7 & 106.4 & 87.1 & 100.2 & 104.5 & 70.8 & 86.3 \\
  Fe I & 5501.46 &  0.96 & -2.95 & 86.8 & 94.8 & 95.8 & 85.6 & 56.5 & 58.2 & 79.8 \\
  Fe I & 5506.78 &  0.99 & -2.80 & 95.0 & 117.2 & 94.9 & 106.4 & 97.7 & 82.3 & 92.5 \\
  Fe I & 5569.62 &  3.42 & -0.54 & ...& ...& ...& 34.0 & ...& 29.5 & 26.9 \\
  Fe I & 5572.84 &  3.40 & -0.31 & ...& 36.2 & ...& ...& 40.1 & ...& 38.4 \\
  Fe I & 5586.76 &  3.37 & -0.14 & 58.9 & 71.3 & 47.3 & 57.9 & 57.5 & ...& 50.0 \\
  Fe I & 5615.64 &  3.33 & -0.14 & 64.0 & 66.6 & 55.8 & 71.8 & 42.9 & 55.0 & 60.9 \\
  Fe I & 6136.62 &  2.45 & -1.40 & 59.1 & 61.7 & 60.2 & ...& ...& 37.9 & 53.4 \\
  Fe I & 6137.69 &  2.59 & -1.40 & 50.1 & 71.8 & 50.4 & 62.4 & 57.7 & 20.3 & 44.1 \\
  Fe I & 6191.56 &  2.43 & -1.60 & 70.5 & 55.6 & 53.2 & 61.1 & 62.9 & 35.1 & 51.2 \\
  Fe I & 6230.72 &  2.56 & -1.28 & ...& ...& ...& 75.7 & ...& ...& 55.4 \\
  Fe I & 6335.33 &  2.20 & -2.23 & 37.5 & 54.2 & 28.0 & 45.3 & 32.5 & ...& 29.0 \\ 
  Fe I & 6393.60 &  2.43 & -1.62 & 47.3 & ...& 50.6 & 60.7 & 50.5 & ...& 48.0 \\
  Fe I & 6421.35 &  2.28 & -2.03 & 39.9 & ...& ...& 32.2 & ...& ...& 35.2 \\
  Fe I & 6430.85 &  2.18 & -2.01 & 64.5 & ...& 54.2 & 42.2 & ...& 31.0 & 45.5 \\
  Fe I & 6677.99 &  2.69 & -1.47 & 55.7 & 69.1 & 38.3 & 49.7 & ...& ...& 37.9 \\
  Ni I & 4714.41 &  3.38 &  0.26 & ...& ...& 48.8 & ...& ...& ...& 20.9 \\
  Ni I & 5137.07 &  1.68 & -1.99 & ...& ...& ...& 32.9 & ...& ...& 21.1 \\
  Ni I & 5476.90 &  1.83 & -0.89 & ...& 95.3 & 78.6 & ...& ...& ...& ...\\
  Sc II & 4415.56 &  0.60 & -0.67 &...&...&...& 68.5 &...& 70.2 & 63.7 \\
  Sc II & 5031.02 &  1.36 & -0.40 &...&...&...& 32.6 &...&...& 31.5 \\
  Sc II & 5526.79 &  1.77 &  0.02 &...& 35.7 & 35.9 & 35.9 & 27.9 &...& 31.4 \\
  Ti II & 4443.79 &  1.08 & -0.72 & 102.0 & 101.4 & 138.6 & 99.8 &...&...& 110.8 \\
  Ti II & 4444.55 &  1.12 & -2.24 &...&...& 61.0 & 26.7 &...&...& 40.2 \\
  Ti II & 4450.48 &  1.08 & -1.50 &...& 123.5 & 102.2 & 67.2 &...&...& 84.0 \\
  Ti II & 4468.51 &  1.13 & -0.60 & 91.7 & 144.5 &...& 96.6 &...&...& 111.0 \\
  Ti II & 4501.27 &  1.12 & -0.77 &...& 76.4 &...& 85.2 & 128.9 & 73.1 & 113.2 \\
  Ti II & 4563.76 &  1.22 & -0.69 &...& 114.2 & 82.8 & 113.2 & 94.7 &...& 112.6 \\
  Ti II & 4571.97 &  1.57 & -0.32 &...&...&...& 88.7 &...&... & 98.6 \\
  Ti II & 4589.96 &  1.24 & -1.62 & 50.2 &...&...& 80.8 &...&... & 66.5 \\
  Ti II & 4657.20 &  1.24 & -2.24 &...&...&...& 50.3 &...&... & 31.2 \\
  Ti II & 4805.08 &  2.06 & -0.96 & 46.9 & 40.2 &...&...&...&... & 39.4 \\
  Ti II & 5154.07 &  1.57 & -1.78 &...& 50.5 &...& 40.9 & 45.3 &...& 37.0 \\
  Ti II & 5185.91 &  1.88 & -1.37 &...&...&...& 37.7 & 66.2 &...& 30.0 \\
  Ti II & 5226.54 &  1.57 & -1.23 & 62.0 &...&...& 76.9 &...&...& 62.4 \\
  Ti II & 5336.79 &  1.58 & -1.59 & 40.9 &...&...& 43.7 & 87.2 &...& 44.5 \\
  Fe II & 4508.29 &  2.86 & -2.31 &...&...&...& 47.3 &...&...& 41.0 \\
  Fe II & 4583.84 &  2.81 & -1.74 &...& 103.8 &...& 75.1 &...&...& 71.8 \\
  Fe II & 4923.93 &  2.89 & -1.21 & 76.3 & 111.3 & 73.2 & 97.2 & 106.3 & 82.5 & 97.4 \\
  Fe II & 5018.43 &  2.89 & -1.23 &...&...& 63.0 &...&...&...& 107.6 \\
  Fe II & 5197.58 &  3.23 & -2.35 & 56.9 &...& 43.2 & 41.9 & 66.2 &...& 27.9 \\
  Fe II & 5234.63 &  3.22 & -2.15 & 51.5 & 43.2 & 37.0 & 37.2 &...& 23.1 & 33.0 \\
  Fe II & 5275.99 &  3.20 & -2.13 & 46.3 & 50.3 & 55.3 & 61.5 &...& 23.1 & 39.8 \\
  Fe II & 5316.62 &  3.15 & -2.02 & 68.2 & 63.5 & 58.5 & 56.2 & 72.1 & 53.6 & 65.3 \\
  Ba II & 4934.08 &  0.00 & -0.15 & 50.7 & 89.0 & 86.5 & 158.3 & 90.3 & 177.9 & 85.7 \\
  Ba II & 5853.69 &  0.60 & -0.91 &...&...&...& 39.1 &...& 69.8 & 8.1 \\
  Ba II & 6141.73 &  0.70 & -0.08 & 42.5 & 45.3 & 27.9 & 83.3 & 25.5 & 119.8 & 38.9 \\
\end{longtable}
}

\begin{thebibliography}{}

\bibitem[Alonso, Arribas, Mart\'{i}nez-Roger
(1999)]{alonso99}Alonso, A., Arribas, S., \& Mart\'{i}nez-Roger,
C. 1999, \aaps, 140, 261

\bibitem[Aldenius et al. (2007)]{aldenius07} Aldenius, M., Tanner,
 J. D., Johansson, S., Lundberg, H.,\& Ryan, S. G.  2007, A\&A 461,
 767-773

\bibitem[Andrievsky et al.(2007)]{andrievsky07} Andrievsky, S.~M.,
  Spite, M., Korotin, S.~A., Spite, F.  , Bonifacio, P., Cayrel, R.,
  Hill, V., \& Fran{\c c}ois, P.\ 2007, \aap, 464, 10 81


\bibitem[Aoki et al. (2005)]{aoki05} Aoki, W., et al.\ 2005, 
\apj, 632, 611 

\bibitem[Aoki et al.(2007a)]{aoki07a} Aoki, W., Beers, T.~C., 
Christlieb, N., Norris, J.~E., Ryan, S.~G., 
\& Tsangarides, S.\ 2007a, \apj, 655, 492 

\bibitem[Aoki et al.(2007b)]{aoki07b} Aoki, W., et al.\ 2007b, 
\apj, 660, 747 

\bibitem[Asplund et al.(2005)]{asplund05} Asplund, M., Grevesse, 
N., \& Sauval, A.~J.\ 2005, ASP Conf.~Ser.~336: Cosmic Abundances as 
Records of Stellar Evolution and Nucleosynthesis, 336, 25 

\bibitem[Battaglia et al.(2008)]{battaglia08} Battaglia, G., Irwin, 
M., Tolstoy, E., Hill, V., Helmi, A., Letarte, B., 
\& Jablonka, P.\ 2008, \mnras, 383, 183 

\bibitem[Beers \& Christlieb(2005)]{beers05} Beers, T.~C., \& 
Christlieb, N.\ 2005, \araa, 43, 531 
 		 
\bibitem[Carney(1983)]{carney83} Carney, B.~W.\ 1983, \aj, 88, 
610 

\bibitem[Castelli \& Kurucz(2003)]{castelli03} Castelli, F., \&
Kurucz, R.~L.\ 2003, Modelling of Stellar Atmospheres, 210, 20

\bibitem[Cayrel et al. (2004)]{cayrel04} Cayrel, R. et al. 2004, A\&A,
416, 1117

\bibitem[Frebel et al.(2009)]{frebel09} Frebel, A., Simon, 
J.~D., Geha, M., \& Willman, B.\ 2009, arXiv:0902.2395 

\bibitem[Fulbright et al.(2004)]{fulbright04} Fulbright, J.~P., 
Rich, R.~M., \& Castro, S.\ 2004, \apj, 612, 447 

\bibitem[Fulbright(2000)]{fulbright00} Fulbright, J.~P.\ 2000, \aj, 
120, 1841 

\bibitem[Helmi et al.(2006)]{helmi06} Helmi, A., et al.\ 2006, 
\apjl, 651, L121 

\bibitem[Hodgkin et al.(2009)]{hodgkin09} Hodgkin, S.~T., Irwin, 
M.~J., Hewett, P.~C., \& Warren, S.~J.\ 2009, \mnras, 394, 675 

\bibitem[Honda et al.(2007)]{honda07} Honda, S., Aoki, W., 
Ishimaru, Y., \& Wanajo, S.\ 2007, \apj, 666, 1189 

\bibitem[Honda et al.(2004)]{honda04} Honda, S., Aoki, W., 
Kajino, T., Ando, H., Beers, T.~C., Izumiura, H., Sadakane, K., \& 
Takada-Hidai, M.\ 2004, \apj, 607, 474 

\bibitem[Irwin \& Lewis(2001)]{irwin01} Irwin, M., \& Lewis, J.\ 2001, New Astronomy Review, 45, 105 

\bibitem[Irwin et al.(2004)]{irwin04} Irwin, M.~J., et al.\ 
2004, \procspie, 5493, 411 

\bibitem[Kim et al.(2002)]{kim02} Kim, Y.-C., Demarque, P., 
Yi, S.~K., \& Alexander, D.~R.\ 2002, \apjs, 143, 499 

\bibitem[Kirby et al.(2008)]{kirby08} Kirby, E.~N., Simon, 
J.~D., Geha, M., Guhathakurta, P., \& Frebel, A.\ 2008, \apjl, 685, L43 

\bibitem[Kobayashi et al.(1998)]{kobayashi98} Kobayashi, C., 
Tsujimoto, T., Nomoto, K., Hachisu, I., \& Kato, M.\ 1998, \apjl, 503, L155 

\bibitem[Fran{\c c}ois et 
al.(2007)]{francois07} Fran{\c c}ois, P., et al.\ 2007, \aap, 476, 935 

\bibitem[Koch et al.(2008)]{koch08} Koch, A., McWilliam, A., 
Grebel, E.~K., Zucker, D.~B., \& Belokurov, V.\ 2008, \apjl, 688, L13 

\bibitem[Kraft(1994)]{kraft94} Kraft, R.~P.\ 1994, \pasp, 106, 
553 

\bibitem[Kurucz (1993)]{kurucz93} Kurucz, R.\ 1993, ATLAS9 
Stellar Atmosphere Programs and 2 km/s grid.~Kurucz CD-ROM No.~13.~ 
Cambridge, Mass.: Smithsonian Astrophysical Observatory, 1993, 13

\bibitem[Lai et al.(2008)]{lai08} Lai, D.~K., Bolte, M., 
Johnson, J.~A., Lucatello, S., Heger, A., 
\& Woosley, S.~E.\ 2008, \apj, 681, 1524 

\bibitem[Lawrence et al.(2007)]{lawrence07} Lawrence, A., et al.\ 
2007, \mnras, 379, 1599 

\bibitem[Mateo(1998)]{mateo98} Mateo, M.~L.\ 1998, \araa, 36, 435 


\bibitem[McWilliam et al.(1995)]{mcwilliam95} McWilliam, A., 
Preston, G.~W., Sneden, C., \& Searle, L.\ 1995, \aj, 109, 2757 

\bibitem[Moity(1983)]{moity83} Moity, J.\ 1983, \aaps, 52, 37 

\bibitem[Noguchi et al. (2002)]{noguchi02} Noguchi, K., et al.\ 
2002, \pasj, 54, 855 

\bibitem[Norris et al.(2008)]{norris08} Norris, J.~E., Gilmore, 
G., Wyse, R.~F.~G., Wilkinson, M.~I., Belokurov, V., Evans, N.~W., 
\& Zucker, D.~B.\ 2008, \apjl, 689, L113 

\bibitem[Pagel\& Patchett(1975)]{pagel75} Pagel, B.~E.~J., \&
  Patchett, B.~E.\ 1975, \mnras, 172, 13

\bibitem[Pickering et al.(2001)]{pickering01} Pickering, J.~C., 
Thorne, A.~P., \& Perez, R.\ 2001, \apjs, 132, 403 

\bibitem[Reddy et al.(2006)]{reddy06} Reddy, B.~E., Lambert, 
D.~L., \& Allende Prieto, C.\ 2006, \mnras, 367, 1329 

\bibitem[Reddy et al.(2003)]{reddy03} Reddy, B.~E., Tomkin, J., 
Lambert, D.~L., \& Allende Prieto, C.\ 2003, \mnras, 340, 304 

\bibitem[Ryabchikova et al.(1994)]{ryabchikova94} Ryabchikova, T.~A., 
Hill, G.~M., Landstreet, J.~D., Piskunov, N., 
\& Sigut, T.~A.~A.\ 1994, \mnras, 267, 697 

\bibitem[Sadakane et al.(2004)]{sadakane04} Sadakane, K., Arimoto, 
N., Ikuta, C., Aoki, W., Jablonka, P., \& Tajitsu, A.\ 2004, \pasj, 56, 
1041 

\bibitem[{Schlegel {et~al.}(1998)Schlegel, Finkbeiner, \&
  Davis}]{schlegel98}
Schlegel, D., Finkbeiner, D., \& Davis, M. 1998, ApJ, 500, 525

\bibitem[Shetrone(1996a)]{shetrone96a} Shetrone, M.~D.\ 1996a, \aj, 
112, 1517 

\bibitem[Shetrone(1996b)]{shetrone96b} Shetrone, M.~D.\ 1996b, \aj, 
112, 2639 

\bibitem[Shetrone et al. (2001)]{shetrone01} Shetrone, M. D., Cote, P.,
Sargent, W. L. W. 2001, ApJ, 548, 592 

\bibitem[Shetrone et al. (2003)]{shetrone03} Shetrone, M., Venn, 
K.~A., Tolstoy, E., Primas, F., Hill, V., \& Kaufer, A.\ 2003, \aj, 125, 
684 

\bibitem[Skrutskie et al.(2006)]{skrutskie06} Skrutskie, M.~F., et
al.\ 2006, \aj, 131, 1163

\bibitem[Spite et 
al.(2005)]{spite05} Spite, M., et al.\ 2005, \aap, 430, 655 

\bibitem[Stephens \& Boesgaard(2002)]{stephens02} Stephens, A., \&
Boesgaard, A.~M.\ 2002, \aj, 123, 1647

\bibitem[Takeda et al.(2003)]{takeda03} Takeda, Y., Zhao, G., 
Takada-Hidai, M., Chen, Y.-Q., Saito, Y.-J., 
\& Zhang, H.-W.\ 2003, Chinese Journal of Astronomy and Astrophysics, 3, 316 

\bibitem[Truran et al.(2002)]{truran02} Truran, J.~W., Cowan, 
J.~J., Pilachowski, C.~A., \& Sneden, C.\ 2002, \pasp, 114, 1293 

\bibitem[Tsujimoto \& Shigeyama (2002)]{tsujimoto02} Tsujimoto, T. \&
Shigeyama, T. 2002, \apj, 571, L93

\bibitem[Venn et al.(2004)]{venn04} Venn, K.~A., Irwin, M., 
Shetrone, M.~D., Tout, C.~A., Hill, V., \& Tolstoy, E.\ 2004, \aj, 128, 
1177 

\bibitem[Woosley \& Weaver(1995)]{woosley95} Woosley, S.~E., \&
  Weaver, T.~A.\ 1995, \apjs, 101, 181

\bibitem[Yoshii et al.(1996)]{yoshii96} Yoshii, Y., Tsujimoto, 
T., \& Nomoto, K.\ 1996, \apj, 462, 266 

\end{thebibliography}
\end{document}